\documentclass[a4paper,11pt]{article}
\usepackage{jheppub}
\usepackage{bm, comment, braket}
\usepackage[x11names]{xcolor}
\usepackage[T1]{fontenc}
\usepackage{comment}
\usepackage{colortbl}
\usepackage{xcolor}
\newcommand{\del}{\partial}
\newcommand{\beq}{\begin{eqnarray}}
\newcommand{\eeq}{\end{eqnarray}}
\newcommand\+{\dagger}

\newcommand{\SU}{\text{SU}}

\usepackage{amsmath}
\allowdisplaybreaks
\usepackage{ulem}
\usepackage{cancel}

\begin{document}

\title{
Holographic QCD Matter: Chiral Soliton Lattices in Strong Magnetic Field
}

\author[a,b]{Markus~A.~G.~Amano,}
\emailAdd{markus(at)oyama-ct.ac.jp}

\author[b,c,d]{Minoru~Eto,}
\emailAdd{meto(at)sci.kj.yamagata-u.ac.jp}

\author[e,c,d]{Muneto~Nitta,}
\emailAdd{nitta(at)phys-h.keio.ac.jp}

\author[f]{
and Shin~Sasaki}
\emailAdd{shin-s(at)kitasato-u.ac.jp}

\affiliation[a]{
Department of General Education, National Institute of Technology, Oyama College, 
777 Ooazanakakuki, Oyama City, Japan
}

\affiliation[b]{
Department of Physics, Yamagata University, 
1-4-12 Kojirakawa-machi, Yamagata, Japan
}

\affiliation[c]{
Research and Education Center for Natural Sciences, Keio University, 
4-1-1 Hiyoshi, Yokohama, Japan
}

\affiliation[d]{
International Institute for Sustainability with Knotted Chiral Meta Matter (WPI-SKCM$^2$), Hiroshima University, 
1-3-2 Kagamiyama, Higashi-Hiroshima, Japan
}

\affiliation[e]{
Department of Physics, Keio University, 
4-1-1 Hiyoshi, Yokohama, Japan
}

\affiliation[f]{
Department of Physics, Kitasato University, 
1-15-1 Kitasato, Sagamihara, Japan
}

%
%
%
%
%
%

\abstract{
We investigate the chiral soliton lattice (CSL) in the framework of
holographic QCD in magnetic field.
Under appropriate boundary conditions for the gauge field and the quark mass
deformation, we demonstrate that the ground state in the gravitational dual of
QCD is given by the CSL in the background magnetic field and the baryon number
density.
In the presence of the background magnetic field, we show that the CSL is
interpreted as a uniformly distributed D4-branes in the holographic setup, where
the chiral soliton is identified with 
a non-self-dual instanton vortex or
a center vortex in the five dimensional bulk gauge theory.
While the baryon numbers are given to chiral solitons as well as Skyrmions due to the different terms in 
the Wess-Zumino-Witten (WZW) term in the chiral perturbation theory, these baryon numbers with different origins are unified in terms of the instanton charge density in five dimensions.
With bulk analysis of the WZW term, we find that the pion decay
constant becomes dependent on the magnetic field.
For the massless pion case,  we obtain  an analytical form  that is in
qualitative agreement with lattice QCD results for strong magnetic fields.
}

\maketitle

\section{Introduction}

Quantum Chromodynamics (QCD) is the fundamental theory describing the strong interaction in terms of quarks and gluons, which has been established by perturbative analysis and lattice simulations. 
QCD under extreme conditions such as high baryon density, strong magnetic fields, or rapid rotation
 has recently received intense attention because of its relevance to the interiors of neutron stars
and heavy-ion collisions \cite{Fukushima:2010bq}. 
Unfortunately, for the aforementioned scenarios, the coupling is large, so perturbative analysis cannot be reliably performed. Traditionally, lattice calculations are done in the non-perturbative regime.
However, the lattice calculations of QCD 
 suffer from the notorious 
sign problem at finite baryon density. 
For large coupling and non-zero chemical potentials, chiral perturbation theory (ChPT) has proven to be quite useful 
  as a low-energy effective theory of QCD.
ChPT has been used to model massless bosons or pions associated with spontaneous chiral symmetry breaking \cite{Scherer:2012xha, Bogner:2009bt}.
 This is a model-independent description determined by only symmetries with UV-determined constants such as the pion's decay constant, $f_{\pi}$, and  mass, $m_{\pi}$.   
Background magnetic fields, $B$, or rotation, $\Omega$,
together with finite chemical potential, $\mu_{\rm B}$,
can be incorporated into ChPT 
by introducing the Wess-Zumino-Witten (WZW) term \cite{Witten:1983tw}. 
The WZW term contains an anomalous coupling of the neutral pion, $\pi^0$, to the magnetic field via the chiral anomaly \cite{Son:2004tq,Son:2007ny}.
The anomalous coupling is mediated through 
the Goldstone-Wilczek current \cite{Goldstone:1981kk,Witten:1983tw}.
It was shown in ref.~\cite{Son:2007ny} that 
the ground state of QCD with two flavors 
in a strong magnetic field $B$ satisfying 
$B \geq B_{\rm CSL} =
    {16\pi m_{\pi}/f_{\pi}^2}{e \mu_{\rm B}}$ 
with the electromagnetic gauge coupling $e$ is not uniform anymore and is modulated   
to form a chiral soliton lattice (CSL), 
 a stack of solitons of the neutral pion, $\pi^0$,
carrying a baryon number density. 
Here, the chiral soliton is a sine-Gordon soliton of the neutral pion field $\pi^0$ 
depending on the direction along the magnetic field.
See also refs.~\cite{Eto:2012qd,Brauner:2016pko,
Brauner:2017uiu,Brauner:2017mui,Brauner:2021sci,Brauner:2023ort} for further study.
QCD under rapid rotations 
also exhibits analogous CSLs composed of  
the $\eta$ or $\eta'$ meson ~\cite{Huang:2017pqe,Nishimura:2020odq,Chen:2021aiq,Eto:2021gyy,Eto:2023tuu,Eto:2023rzd}.
Furthermore, the mixture of the $\pi_0$ and $\eta'$ solitons leads to the formation of quasicrystals \cite{Qiu:2023guy}.
These results are obtained at
zero temperature, 
but thermal fluctuations enhance 
the stability of CSLs~\cite{Brauner:2017uiu, Brauner:2017mui, Brauner:2021sci, Brauner:2023ort}. 
The formation of CSLs were studied 
as nucleation through quantum tunneling 
\cite{Eto:2022lhu,Higaki:2022gnw} and 
dynamical formation \cite{Eto:2025ebz}.
It has been shown that for large densities in strong magnetic fields \cite{Eto:2023lyo,Eto:2023wul,Amari:2024fbo} or under rapid rotation \cite{Eto:2023tuu} the CSL phase undergoes a phase transition into the so-called domain-wall Skyrmion phase.
This phase is characterized by the spontaneous formation of Skyrmions \cite{Skyrme:1961vq, Skyrme:1962vh} 
on top of chiral solitons.
These domain-wall Skrymions have a baryon number of two and are bosons \cite{Amari:2024mip}.
Though it has yet to be shown, it is expected that large-$N_{\rm C}$ QCD admits
stable phase mixtures between the domain-wall Skyrmion phase and a Skyrmion crystal phase \cite{Klebanov:1985qi}.
For more on this topic, see  refs.~\cite{Kawaguchi:2018fpi,Chen:2021vou,Chen:2023jbq, Amari:2025twm}.
There are also other studies that look into Abrikosov's vortex lattices/baryon crystals \cite{Evans:2022hwr, Evans:2023hms, Kaplunovsky:2015zsa} and a vortex Skyrmion phase with a non-zero isospin chemical potential \cite{Qiu:2024zpg} (see also ref.~\cite{Gronli:2022cri}).
In order to clarify the stability of CSLs in the non-perturbative regime, researchers have investigated CSL ground states in QCD-like theories such as $\SU(2)$ QCD and vector-like gauge theories, which can circumvent the sign problem in lattice gauge theory at finite baryon density~\cite{Brauner:2019rjg,Brauner:2019aid}. Supersymmetric QCD has also been studied in this context~\cite{Nitta:2024xcu}.

Our motivation is to clarify the fate of CSL phases at the non-perturbative
regime of QCD. It is well known that the first principle approach, lattice QCD,
has the aforementioned sign problem at finite baryon density. We thus consider
another approach for strongly coupled dynamics, namely holographic QCD or the
gauge/gravity correspondence based on string theory. The gauge/gravity
correspondence has provided a novel framework to study the strongly coupled
gauge theories through the weakly coupled dual gravitational descriptions. In
particular, the Sakai-Sugimoto (SS) model \cite{Sakai:2004cn, Sakai:2005yt},
based on the D4-D8-$\overline{\text{D8}}$-brane configuration in type IIA string theory, offers a
holographic microscopic description of the low-energy QCD. This model
incorporates key features of QCD such as confinement, spontaneous breaking of
the chiral symmetry, and also predicts the existence baryons and mesons. One of
the notable advantages of this model is its geometric realization of the chiral
symmetry, including its spontaneous breaking. This arises naturally from the
configuration of the probe D8-branes in the background geometry of the
D4-branes. In addition, the model successfully, qualitatively reproduces the
meson spectrum, the internal structure of baryons, and various anomalous
processes via the WZW term \cite{Hata:2007mb, Hong:2007kx, Hashimoto:2008zw}.
These features make this holographic model a powerful tool to phenomenologically
explore the strongly coupled, non-perturbative regime of QCD-like theories
beyond the reach of conventional methods. Holographic QCD under external
magnetic fields and/or a finite baryon chemical potential has been explored in
refs.~\cite{Rebhan:2008ur, Preis:2011sp, Evans:2022hwr, Callebaut:2011ab,
Burikham:2011rg, Fukushima:2013zga,Bartolini:2023wis, Bartolini:2023eam,
Kovensky:2023mye}.

In this paper, we investigate the CSL within the framework
of holographic QCD, with a particular focus on their realization as
specific D-brane configurations and their behavior in the bulk geometry. 
The SS model, formulated in terms of D8-branes embedded in a
D4-brane background allows us to find the explicit CSL configuration.
Within this setup, we explore how the spatially modulated patterns of
the chiral condensate, such as CSL, can be represented by nontrivial
configurations of D-branes. 
In the presence of the background magnetic field, we show that the CSL is interpreted as a uniformly distributed D4-branes in the holographic setup, 
where the chiral soliton is identified with 
 a non-self-dual instanton vortex or 
a center vortex in the five dimensional bulk gauge theory. 
While chiral solitons and Skyrmions carry the baryon number due to the different terms of the WZW term in the framework of the ChPT, these baryon numbers are understood in a unified way as the instanton charge in five dimensions.
This perspective also enables us to study the bulk realization of the
CSL from a string-theoretic viewpoint, going beyond effective
descriptions limited to the boundary field theory. 
With bulk analysis of the WZW term, we find that the pion decay
constant dependents on the magnetic field.
For the massless pion case, we obtain an analytical form that is in qualitative
agreement with lattice QCD results 
\cite{Shushpanov:1997sf, Simonov:2015xta} 
in strong magnetic fields. We seek to shed light on the
non-perturbative dynamics underlying the spatially modulated chiral phases.
See \cite{Thompson:2008qw, Bergman:2008qv} for related works on the background magnetic field in the SS model.

The curved background of the D4-branes causes the D8 and $\overline{\text{D8}}$-branes to join
at the tip of the geometry. In the field theory, this manifests as chiral
symmetry breaking that is achieved without introducing massive quarks.
This is different from ChPT, in which chiral symmetry is broken with massive
quarks, while the SS model quarks are all massless. Yet, the stability of the
CSL vacuum requires a non-zero quark mass, which induces the mass term in ChPT.
Because of this, introducing a pion mass term into the SS model is crucial to
inducing a CSL phase. However, due to the particular brane configuration of the
SS model, introducing a pion mass has historically been challenging. At present,
there are two major methods to introduce a pion mass. The first is through the
introduction of a non-local Wilson line
operator~\cite{Seki:2013nta,Seki:2012tt,Aharony:2008an,McNees:2008km,Kovensky:2019bih}.
This method adds a non-local action term to the bulk dynamics, but has been
useful in incorporating the effects of massive quarks into the SS model. The
other method involves introducing a tachyonic scalar field, generalizing the SS
model~\cite{Bergman:2007pm,Seki:2010ma,Dhar:2008um,Iatrakis:2010jb}.

The organization of this paper is as follows.
In the next section, we give a brief introduction to the holographic
QCD model \cite{Sakai:2004cn, Sakai:2005yt} that we employ in this paper.
We also introduce the quark mass deformation and the associated brane
configurations \cite{Hashimoto:2008sr, Aharony:2008an}.
We show that the CSL appears as a ground state in the gravity dual with
the appropriate boundary conditions of gauge fields.
In section \ref{sec:brane_CSL}, we discuss a brane interpretation of the
CSL in the holographic model.
We show that the CSL induces the non-zero instanton density $\text{Tr}
[F \wedge F]$ which plays a role of the source term for D4-branes.
We show that this term indeed gives the non-zero baryon number
density discussed in the literature.
In section \ref{sec:bulk_CSL}, we study the bulk configuration of the
CSL.
Section \ref{sec:conclusion} is devoted to conclusion and discussions. 
A derivation of the WZW term in our holographic set up is summarized in appendix \ref{sec:Z_calculation}.
Detail calculations are found in appendix \ref{app:solution_m0Bn0}.
The bulk mass term is proposed in appendix
\ref{app:mass_action_variation}.

\section{Chiral soliton lattices in holographic QCD}
\label{sec:CSL_hQCD}
In this section, we briefly introduce the holographic QCD model that we
discuss in this paper.
We will find that the CSL appears as a ground state in the gravity dual 
when we consider appropriate boundary conditions on the gauge fields in
the D8-branes.

\subsection{The model}
We employ the holographic setup known as the SS  model \cite{Sakai:2004cn,Sakai:2005yt}.
The $N_c$ D4-branes and $N_f \ll N_c$ (anti-)D8-branes are placed as in
table \ref{tb:holographic_setup}.
\begin{table}[t]
\begin{center}
\begin{tabular}{c|c|c|c|c|c|c|c|c|c|c}
 & 0 & 1 & 2 & 3 & (4) & 5 & 6 & 7 & 8 & 9 \\
\hline
$N_c$ D4 & $\circ$ & $\circ$ & $\circ$ & $\circ$ & $\circ$ &  &  &  &  &  \\
\hline
$N_f$ D8/$\overline{\text{D8}}$ & $\circ$ & $\circ$ & $\circ$ & $\circ$ &  & $\circ$ & $\circ$ & $\circ$ & $\circ$ & $\circ$ \\
\hline
$N'$ D6 & $\circ$ & $\circ$ & $\circ$ & $\circ$ & $\circ$ &  & $\circ$ & $\circ$ &  &  
\end{tabular}
\end{center}
\caption{
The brane configuration of the holographic QCD model
 \cite{Sakai:2004cn,Sakai:2005yt,Hashimoto:2008sr}.
The circle $\circ$ indicates the worldvolume directions of the branes.
The D4-branes are located at $x^5 = \cdots = x^9 = 0$ while 
the D8- and $\overline{\text{D8}}$-branes are located at the antipodal
 points $x^4 = 0$ and $\pi M_{\text{KK}}^{-1}$ in $S^1$, respectively.
The D6-branes for the mass deformation are located at $x^5 \not= 0$, $x^8 = x^9 = 0$.
}
\label{tb:holographic_setup}
\end{table}
The $x^4$-direction is compactified on $S^1$ with the Kaluza-Klein (KK)
energy scale $M_{\text{KK}}$.
The $U(N_c)$ gauge fields and the massless $U(N_f)_L \times U(N_f)_R$
fundamental quarks $q_L^f, q_R^f$ live on the D4-brane worldvolume.
Although the D8-$\overline{\text{D8}}$ open strings give rise to a tachyon $T$,
it becomes (positively) massive and cane be safely neglected when the D8-branes and the
$\overline{\text{D8}}$-branes are far enough separated as shown in table
\ref{tb:holographic_setup}.
Supersymmetry (SUSY) is broken due to the anti-periodic conditions on
fermions.
Due to the absence of SUSY, the adjoint scalar fields $\phi^{i} (i=5,\cdots,9)$
become massive via loop effects.
The trace part of the gauge field component $A_{4}$ and $\phi^i$ remain massless but are ignored in the IR since they couple to
other parts through irrelevant operators.
Then, below the KK scale $M_{\text{KK}}$, the field theory on the D4-branes is
a non-SUSY $U(N_c)$ QCD in four dimensions.
Note that the quarks are massless in the model
\cite{Sakai:2004cn,Sakai:2005yt} and there is an exact $U(N_f)_L \times
U(N_f)_R$ chiral symmetry.

In order to introduce a non-zero mass for the quarks $q_L^f, q_R^f$, 
we introduce extra $N' \ll N_c$ D6-branes (table
\ref{tb:holographic_setup}).
These extra $N'$ D6-branes induce a non-zero mass for the quarks via worldsheet instanton
effects \cite{Hashimoto:2008sr, Aharony:2008an}.
From the field theory viewpoint, the condensation of additional quarks
$Q_L, Q_R$ that emerge from the open strings attached to the D6-branes induces the mass term for the quarks $q_L, q_R$.
Thus, we have $U(N_c)$ QCD with massive $N_f$ quarks on D4-branes
where the chiral symmetry is explicitly broken due to the non-zero mass.

In the holographic dual description, the dynamics are governed by the
weakly coupled, gauge-invariant meson operators.
The effective theory of mesons is given by the worldvolume theory of the
D8-branes.
In the large $N_c$ limit ($N_c \gg N_f, N'$), the D4-branes are replaced by the classical background solution, while the D8 ($\overline{\text{D8}}$) and D6-branes are treated as probes and do not backreact on each other's embedding geometry.
The embedding of the D6-branes is discussed in ref.~\cite{Hashimoto:2008sr, Aharony:2008an}.
The latter do not modify the background geometry for the $N_c$ D4-branes in the
large-$N_c$ limit.
The supergravity solution for the D4-brane background is given by
\begin{align}
&
ds^2 = 
\left( 
\frac{U}{R} 
\right)^{\frac{3}{2}} 
\Big(
\eta_{\mu \nu} dx^{\mu} dx^{\nu}
+
f (U) d \tau^2
\Big)
+
\left(
\frac{R}{U}
\right)^{\frac{3}{2}}
\Bigg(
\frac{dU^2}{f (U)} + U^2 d \Omega^2_4
\Bigg),
\notag \\
&
e^{\phi} = g_s \left( \frac{U}{R} \right)^{\frac{3}{4}},
\qquad
F_4 = d C_3 = \frac{2 \pi N_c}{V_4} \epsilon_4,
\qquad
f (U) = 1 - \frac{U_{\text{KK}}^3}{U^3},
\label{eq:D4_sol}
\end{align}
where $\phi$, $C_3$ are the dilaton and the RR 3-form.
Here $x^{\mu} \ (\mu = 0,1,2,3)$ and $x^4 = \tau \sim \tau + 2 \pi M_{\text{KK}}^{-1}$ are coordinates along the
D4-brane extension and $U \ge U_{\text{KK}}$ corresponds to the radial coordinate transverse to
the D4-branes in the $(x^5,x^6,x^7,x^8,x^9)$ directions.
The quantities $d \Omega_4^2$, $\epsilon_4$, $V_4 = \frac{8 \pi^2}{3}$
are the line element, the volume form, and the volume of a unit $S^4$
surrounding D4-branes, respectively.
We have introduced the constants $R^3 = \pi g_s N_c l_s^3, M_{\text{KK}} = \frac{3}{2}
\frac{U_{\text{KK}}^{\frac{1}{2}}}{R^{\frac{3}{2}}}, U_{\text{KK}}$ where 
$g_s$ and $\alpha'$ are the string coupling constant and the slope parameter.
The Yang-Mills and the 't Hooft couplings are given by $g^2_{\text{YM}} = 2
\pi M_{\text{KK}} g_s l_s$, $\lambda = g_{\text{YM}}^2 N_c$.
Note that we always consider the decoupling limit $\alpha' \to 0$,
$g_s^2 N_c = \text{fixed}$ in the holographic setup.
In the D4-brane background, the D8-$\overline{\text{D8}}$ pair 
smoothly connects to form a curved configuration of the D8-branes.
This means that the $U(N_f)_L \times U(N_f)_R$ chiral symmetry is
spontaneously broken down to its diagonal subgroup $U(N_f)_V$ in the IR
(in the massless case).
The connected curved D8-branes are extended along $x^{\mu} \ (\mu = 0,1,2,3)$, the $S^4$
directions, and one of the directions in the $(U,\tau)$-plane.
It is convenient to introduce the coordinates
\begin{align}
(y,z) =
\Big(
\sqrt{U^3 - U_{\text{KK}}^3} 
\cos (\tau M_{\text{KK}}), 
\sqrt{U^3 - U_{\text{KK}}^3} 
\sin (\tau M_{\text{KK}})
\Big).
\end{align}
In these coordinates, the D8-branes are located at $y = 0$ and they are extended along the $(x^{\mu}, z)$- and
the $S^4$-directions.

The action on the gravity side of the D4/D8/$\overline{\text{D8}}$ system is given by that of the
D8-branes on the D4-brane background \eqref{eq:D4_sol}.
At order $\mathcal{O}(\alpha'^2)$, the Dirac-Born-Infeld and the Chern-Simons actions for the D8-branes are given by
\begin{align}
S_{\text{D8}}^{\text{DBI}} =& \  - \kappa \int \! d^4 x dz \,
\text{Tr}
\left[
\frac{1}{2} K^{- \frac{1}{3}} F^2_{\mu \nu} + K F^2_{\mu z}
\right],
\label{eq:DBI}
\\
S_{\text{D8}}^{\text{CS}} =& \ 
\frac{N_c}{24 \pi^2} \int_{M^4 \times \mathbb{R}} \!\! \omega_5 (A),
\label{eq:CS}
\\
\kappa =& \ \frac{\lambda N_c}{2 16 \pi^3} = \frac{\pi}{4}f_\pi^2, \quad K(z) = 1 + \frac{z^2}{U_{\text{KK}}^2}  ,
\label{eq:K}
\end{align}
where $M^4 \times \mathbb{R}$ is spanned by $(x^{\mu},z)$ and we have performed the dimensional reduction along the $S^4$-direction; $A_{6,7,8,9}$ have been set to zero.
The adjoint scalar fields on the D8-branes are also neglected for simplicity.
We use the convention that the gauge fields are hermitian.
The field strength is defined by $F_{\mu \nu} = \del_{\mu} A_{\nu} -
\del_{\nu} A_{\mu} + i [A_{\mu}, A_{\nu}]$.
Here, $\omega_5 (A)$ is the Chern-Simons five-form for the gauge field $A = A_{\mu} d
x^{\mu} + A_z dz$. This is given by
\begin{align}
\omega_5 (A) = \text{Tr}
\Big(
A F^2 - \frac{i}{2} A^3 F - \frac{1}{10} A^5
\Big).
\end{align}

In order to obtain the four-dimensional theory, one can decompose the bulk
fields in terms of a complete set of orthonormal functions for $ z $:
\begin{align}
A_{\mu} (x,z) =& \ \sum_{n=1}^{\infty} B^{(n)}_{\mu} (x) \psi_n (z),
\notag \\
A_z (x,z) =& \ \varphi^{(0)} (x) \phi_0 (z) + \sum_{n=1}^{\infty}
 \varphi^{(n)} (x) \phi_n (z),
\end{align}
where the functions $\psi_n (z), \phi_n (z)$ are normalized as
\begin{align}
&
\kappa \int \! dz \, K^{-1/3} \psi_n \psi_m = \delta_{mn},
\notag \\
&
\kappa \int \! dz \, K \phi_n \phi_m = \delta_{mn}.
\end{align}
Using these orthogonal functions, the $z$-direction is integrated out
and the effective action for $\varphi^{(0)} (x)$ (pion) and
$B^{(n)}_{\mu} (x)$ (vector mesons) is obtained.

By weakly gauging the chiral symmetry $U(N_f)_L \times U(N_f)_R$, 
the background external gauge fields $A_{L \mu} (x)$ and $A_{R \mu} (x)$
are introduced. 
These are defined by the asymptotic values of the gauge field $A_{\mu} (x,z)$ on
the D8-branes:
\begin{align}
\lim_{z \to + \infty} A_{\mu} (x,z) = A_{L \mu} (x),
\qquad
\lim_{z \to - \infty} A_{\mu} (x,z) = A_{R \mu} (x).
\end{align}
Then the mode expansion is given by
\begin{align}
A_{\mu} (x,z) = A_{L \mu} (x) \psi_+ (z) + A_{R \mu} (x) \psi_- (z) +
 \sum_{n=1}^{\infty} B^{(n)}_{\mu} (x) \psi_n (z),
\end{align}
where the functions $\psi_{\pm} (z)$ are defined by
\begin{align}
\psi_{\pm} (z) = \frac{1}{2} \Big( 1 \pm \psi_0 (z) \Big),
\qquad
\psi_0 (z) = \frac{2}{\pi} \arctan 
\left( \frac{z}{U_{\text{KK}}} \right).
\end{align}
It is convenient to employ a gauge where $A_z = 0$.
In this gauge, we instead have
\begin{align}
A_{\mu} (x,z) = 
A^{\xi_+}_{L \mu} (x) \psi_+ (z)
+
A^{\xi_-}_{R \mu} (x) \psi_- (z)
+
\sum_{n=1}^{\infty} B^{(n)}_{\mu} (x) \psi_n (z),
\label{eq:field_expansion}
\end{align}
where
\begin{align}
A^{\xi_+}_{L \mu} (x) =& \ \xi_+ (x) A_{L\mu} (x) \xi_+^{-1} (x) 
-i \,
\xi_+ (x) \del_{\mu} \xi^{-1}_+ (x),
\notag \\
A^{\xi_-}_{R \mu} (x) =& \ \xi_- (x) A_{R \mu} (x) \xi_-^{-1} (x) 
-i \,
\xi_- (x) \del_{\mu} \xi^{-1}_- (x).
\label{eq:field_expansion2}
\end{align}
The fields $\xi_{\pm} (x)$ carry the pion degrees of freedom.
The boundary conditions are 
\begin{align}
\lim_{z \to \infty} A_{\mu} (x,z) = A^{\xi_+}_{L \mu} (x),
\qquad
\lim_{z \to - \infty} A_{\mu} (x,z) = A^{\xi_-}_{R \mu} (x).
\end{align}
The gauge transformation is 
\begin{align}
A_{L \mu} \to& \ g_+ A_{L \mu} g^{-1}_+ 
-i 
g_+ \del_{\mu} g_+^{-1},
\notag \\
A_{R \mu} \to& \ g_- A_{R \mu} g^{-1}_- 
-i
g_- \del_{\mu} g_-^{-1},
\notag \\
\xi_{\pm} \to& \ h \xi_{\pm} g^{-1}_{\pm},
\notag \\
B^{(n)}_{\mu} \to& \ h B^{(n)}_{\mu} h^{-1}.
\end{align}
Here $(g_+(x), g_- (x)) \in U(N_f)_L \times U(N_f)_R$ and $h(x) \in
U(N_f)$.
The $U(N_f)$-valued pion field 
$\Pi = \Pi^a T^a$ is defined by 
\begin{align}
\xi^{-1}_+ (x) \xi_- (x) = U(x) = e^{2 i \Pi (x)/f_{\pi}}
\end{align}
where $f_{\pi}$ is the pion decay constant whose mass
dimension is $[f_{\pi}] = +1$. 
The $U(N_f)$ generators $T^a$ are normalized as $\text{Tr} [T^a T^b] = \frac{1}{2} \delta^{ab}$.
In the following, we employ the gauge where $\xi_+^{-1} = \xi_- = e^{i \Pi/f_{\pi}}$.
The $U(N_f)_L \times U(N_f)_R$ gauge transformation of the pion field is given by
\begin{align}
 U (x) \to g_+ (x) U (x) g^{-1}_-.
\end{align}
The Chern-Simons term of the D8-brane action in the $A_z = 0$ gauge is given by
\begin{align}
S_{\text{D8}}^{\text{CS}} =& \ 
- \frac{N_c}{24 \pi^2} \int_{M^4} 
\!
\Big(
\alpha_4 (d \xi^{-1}_+ \xi_+, A_L)
-
\alpha_4 (d \xi^{-1}_- \xi_-, A_R )
\Big)
\notag \\
& \ + 
\frac{N_c}{24 \pi^2} \int_{M^4 \times \mathbb{R}}
\!
\Bigg(
\omega_5 (A) - \frac{1}{10} \text{Tr} (g d g^{-1})^5
\Bigg),
\end{align}
where we have defined
\begin{align}
\alpha_4 (V,A) =
\frac{1}{2} 
\text{Tr}
\Bigg(
V (A d A + d A A + 
i A^3)
-
\frac{1}{2}
V A V A  
+i V^3 A
\Bigg).
\end{align}

\subsection{Chiral Lagrangian in gravity side}
\label{sec:chiral_model}
In the $A_z = 0$ gauge, we have the explicit field expansion \eqref{eq:field_expansion}.
We set the vector mesons $B_{\mu}^{(n)} = 0$ in the following.
To leading order, the effective action in the gravity side in our setup is
given by $S_{\text{grav.}} = S^{\text{DBI}}_{\text{D8}} +
S^{\text{CS}}_{\text{D8}} + S_{\text{mass}}$, where each part is calculated as 
\begin{align}
S^{\text{DBI}}_{\text{D8}} =& \ \int \! d^4 x \, 
\Bigg(
\frac{f_{\pi}^2}{4} \text{Tr} 
\Big(
U^{-1} 
D_{\mu} U
\Big)^2
+
\frac{1}{32 e^2_S} 
\text{Tr}
\Big[
U^{-1} D_{\mu} U
, 
U^{-1} D_{\nu} U
\Big]^2
\Bigg)
,
\\
\label{eq:action_pionmass}
S_{\text{mass}} =& \ 
\int \! d^4 x \, 
\frac{m_\pi^2 f_{\pi}^2}{4}
\text{Tr}
\Big(
U + U^{-1} - 2
\Big),
\end{align}
and
\begin{align}
S^{\text{CS}}_{\text{D8}} =& \ 
\frac{N_c}{48 \pi^2} \int_{M^4} Z
+ \frac{N_c}{240 \pi^2} \int_{M^4 \times \mathbb{R}} \text{Tr} (g d g^{-1})^5,
\notag \\
Z =& \ 
\text{Tr}
\Bigg[
\Big\{
\Big(
A_R d A_R + d A_R A_R + i A_R^3
\Big)
\Big(
- U^{-1} A_L U + i U^{-1} d U
\Big)
- \text{p.c.}
\Big\}
\notag \\
& 
+ 
\Big(
i d A_R d U^{-1} A_L U - \text{p.c.}
\Big)
+ 
\Big(
A_R (d U^{-1} U)^3 - \text{p.c.}
\Big)
\notag \\
&
+ 
\Big(
\frac{i}{2} (A_R d U^{-1} U)^2 - \text{p.c.}
\Big)
+ 
\Big(
i U A_R U^{-1} A_L d U d U^{-1} - \text{p.c.}
\Big)
\notag \\
&
+
\Big(
A_R d U^{-1} U A_R U^{-1} A_L U - \text{p.c.}
\Big)
- \frac{i}{2}
\Big(
A_R U^{-1} A_L U
\Big)^2
\Bigg].
\label{eq:D8CS_term}
\end{align}
The $S_{\text{mass}}$ action can be constructed by the insertion of non-local operators or non-renormalizable tachyonic scalar fields~\cite{Seki:2013nta, Seki:2012tt, Aharony:2008an, McNees:2008km, Kovensky:2019bih, Niarchos:2010ki}.

Here we have performed the $z$-integration and 
$f_{\pi}, e_S, m_\pi$ are constants.
The symbol p.c.~means the terms obtained by the replacements 
$A_L \leftrightarrow A_R$ and $U \leftrightarrow U^{-1}$.
The covariant derivative is defined by 
$D_{\mu} U = \del_{\mu} U + i A_{L \mu} U - i U A_{R \mu}$.
We have ignored the divergent kinetic term for the background gauge field in $S_{\text{grav.}}$. 
Although this term is divergent, it does not depend on the dynamical fields, so we can safely neglect it~\cite{Sakai:2005yt}.
The action $S_{\text{grav.}}$ is nothing but the one for the chiral
perturbation theory of mesons. 

The electromagnetic gauge group $U(1)_{\text{em}}$ in the chiral perturbation theory is a part of
the gauged chiral symmetry, $SU(N_f)_L \times SU(N_f)_R \to SU(N_f)_V \supset
U(1)_{\text{em}}$.
For example, in the $N_f = 3$ case, we have
$A_L = A_R = Q A_{\text{em}}$ where $Q = \frac{e}{3} \text{diag} (2,-1,-1)$.
However, the baryon number gauge group $U(1)_{\text{B}}$ is not a part of the $SU(N_f)_L \times
SU(N_f)_R$ chiral symmetry~\cite{Witten:1983tw}.
We consider the diagonal subgroup of the chiral symmetry: $U(N_f)_L \times U(N_f)_R \to
U(N_f)_V = SU(N_f) \times U(1)$. The electromagnetic gauge group is a Cartan
subgroup $SU(N_f) \supset U(1)_{\text{em}}$, and the baryon number group $U(1)_B$
is identified with the extra $U(1)$ in this setup.

We focus on $N_f = 2$ as the simplest but physically relevant case,
and the pion fields $\phi^a$ are introduced as 
$U = e^{2 i f_{\pi}^{-1} \frac{\tau^a}{2} \phi^a}$. Here $\tau^a \, (a=1,2,3)$ are the Pauli matrices.
In the following, we focus only on the neutral pion $U = e^{i f^{-1}_{\pi} \tau^3 \phi}$.
The background gauge fields are introduced in the following way.
Assuming the diagonal gauge field $A_L = A_R = A$
, we have 
$A_{\mu} = A_{\text{em} \mu} Q + A_{B \mu}
$ where $Q = \frac{\tau^3}{2} + \frac{1}{6} \mathbf{1}_2$.
Following this, we will omit $\frac{1}{6} \mathbf{1}_2$ since it does not contribute to the energy of the neutral pion $\pi^0$.
Then, in the $SU(N_f = 2)$ sector, 
we turn on $
A_{\text{em} \mu}^3 \frac{\tau^3}{2}$, 
and 
assume $\del_0 A_{\text{em}  i} = A_{\text{em}  0} = 0$. 
In the $U(1)_B$ sector, 
we turn on only $A_{B 0}$ as a constant value.

We also consider the static pion field configuration $\del_0 U = 0$.
In this setup, the relevant contribution in eq.~\eqref{eq:D8CS_term}
is only the term $-i \text{Tr} \big[ F A (U d U^{-1} + d U^{-1} U) \big]$
(see appendix \ref{sec:Z_calculation}).

Then the effective action of the neutral pion $\phi$ on the gravity side is given by
\begin{align}
S_{\text{grav.}} =& \ 
\int \! d^4 x \, 
\text{Tr}
\Bigg\{
\frac{f_{\pi}^2}{4} 
\Big(
U^{-1} D_{\mu} U
\Big)^2
+
\frac{1}{32 e^2_S}
\Big[
U^{-1} D_{\mu} U, \, 
U^{-1} D_{\nu} U
\Big]^2
\notag \\
& \hspace{2cm}
+
\frac{m_\pi^2 f_{\pi}^2}{4}
(U + U^{-1} - 2)
\Bigg\}
+ 
\frac{N_c}{48 \pi^2}
\int \! Z|_{\phi}
\notag \\
=& \ 
\int \! d^4 x \,
\Bigg\{
- \frac{1}{2} (\del_{\mu} \phi)^2
+ f_{\pi}^2 m_\pi^2 (\cos \left(f_{\pi}^{-1} \phi\right) - 1)
+ 
\frac{\mu_B}{4\pi^2 f_\pi}
B_i \del_i \phi
\Bigg\}
\label{eq:gravity_pion}
\end{align}
where we have used 
$
U = e^{i f_{\pi}^{-1} \tau^3 \phi} 
= \mathbf{1}_2 \cos (f_{\pi}^{-1} \phi)
+ i \tau^3 \sin (f_{\pi}^{-1} \phi)
$
and defined 
$\varepsilon^{0ijk} F_{\text{em} \, ij} = - 2 B_k$, $A_{B \, 0} = \mu_B$.
The action \eqref{eq:gravity_pion} is nothing but that of the chiral
sine-Gordon model in a background magnetic field $B_i$ and baryon chemical potential $\mu_B$~\cite{Son:2007ny}.
Assuming that $B_i = (0,0,B_3)$ and $\mu_B$ are constants, the last term in eq.~\eqref{eq:gravity_pion} becomes a total derivative.
This term does not contribute to the equation of motion but does contribute to the
energy.

Indeed, for the massive case with $m_\pi \neq 0$, in a large background $\mu_B |\vec{B}|$,
the periodic kink solution becomes energetically favorable and is given by
\begin{align}
\phi (x^3) = f_{\pi} 
\Bigg[
\pi \pm 2 
\,
\text{am} 
\Big(
m_\pi k^{-1} x^3, k
\Big)
\Bigg]\,.
\label{eq:CSL}
\end{align}
$0 \le k \le 1$ is an elliptic modulus parameter, and the period is $\ell = 2k {\cal K}(k)/m_\pi$ with ${\cal K}(k)$ being the complete elliptic integral of the first kind.
The solution with the upper (lower) sign for $ B_3 > 0 $ ($ B_3 < 0 $) yields lower energy than that of the trivial QCD vacuum, $ \phi = 0 $~\cite{Brauner:2016pko}.
This is the CSL.
The tension of the CSL for one period reads 
\begin{eqnarray}
\sigma = 4 m_\pi f_\pi^2 \left\{
\frac{2}{k}E(k) + \left(\frac{1}{k} - k\right){\cal K}(k)
\right\} \mp \frac{\mu_B B_3}{2\pi} \,,
\end{eqnarray}
with $E(k)$ being the complete elliptic integral of the second kind.
The mean tension $\sigma/\ell$ is minimized for $k$ satisfying
\begin{eqnarray}
\frac{E(k)}{k} = \frac{\mu_B |B_3|}{16 \pi m_\pi f_\pi^2}\,.
\end{eqnarray}
This determines $k$ for a given $B_3$, and the mathematical equation $E(k)/k \ge 1$~\cite{Brauner:2017mui} gives the physical condition for the CSL to be the ground state
\begin{eqnarray}
\label{eq:CSL_inequality}
\mu_B|B_3| \ge 16\pi m_\pi f_\pi^2.
\end{eqnarray}
Note that in the $k \to 1$ limit, corresponding to $|B_3| \to \frac{16\pi m_\pi f_\pi^2}{\mu_B}$, the solution \eqref{eq:CSL} becomes the sine-Gordon kink.

For the massless case $m_\pi=0$, the CSL solution is linear and the corresponding energy density is a negative constant if $B_3 \neq 0$ as
\begin{eqnarray}
\phi(x^3) = \frac{\mu_B B_3}{4\pi^2f_\pi} x^3\,,\quad
{\cal E} = - \frac{\mu_B^2 B_3^2}{32\pi^4 f_\pi^2}\,.
\end{eqnarray}
The massless solution can be obtained from the massive solution as the limit $m_\pi \to 0$ with the ratio 
$2 f_\pi m_\pi / k = \mu_B |B_3|/4\pi^2 f_\pi$
being fixed. The period in the same limit becomes 
$\ell \to 2(k/m_\pi){\cal K}(0) =  8\pi^3 f_\pi^2/\mu_B |B_3|$
which is consistent with the linear slope of the massless solution.
We then find that the ground state of the gravity dual of QCD in large backgrounds is given by the CSL.

\section{Brane interpretation of chiral soliton lattices}
\label{sec:brane_CSL}
In this section, we study a brane interpretation of the CSL in the
holographic setup.
As we have demonstrated, the pion field that forms the CSL and the background field result in the
excitation of the gauge field in the effective theory of the D8-branes.
In this viewpoint, the gauge field induces nontrivial sources of the RR fields on the worldvolume of the D8-branes.
The field configurations are then interpreted as D-branes that intersect with the D8-branes.
In the following, we study how the CSL together with the background field induce RR couplings 
and discuss a brane interpretation of them.


\subsection{Chiral soliton lattices as a brane configuration}\label{sec:CSL-brane}
\begin{figure}[t]
\begin{center}
\includegraphics[scale=0.85]{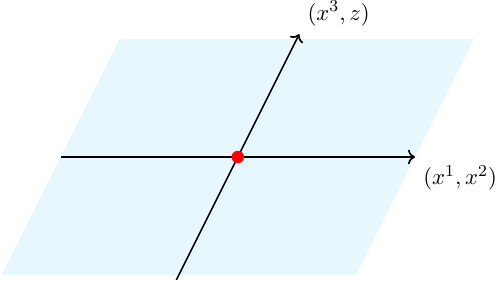}
\includegraphics[scale=0.85]{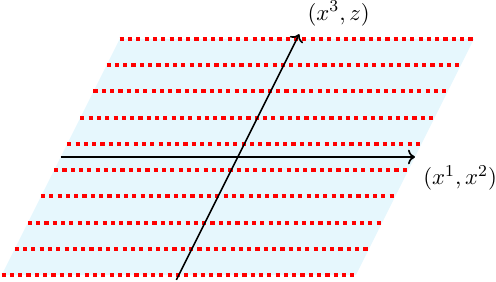}
\end{center}
\caption{
A schematic picture of the brane configuration for the localized
 D4-brane (the red dot in the left figure) and the partially dissolved, periodic 
 D4-branes (the dotted lines in the right figure).
The former is identified with a Skyrmion while the latter corresponds to the CSL in the background field.
The blue rectangle is the D8-branes that fill the whole $(x^1,x^2,x^3, z)$-plane.
The (partially) dissolved D4-branes are periodically localized in the $(x^3, z)$-plane but uniformly
 distributed along the $(x^1,x^2)$-directions.
Note that the worldvolume of the D4-branes is extended along the $(x^0,
 x^6,x^7,x^8,x^9)$-directions, which are omitted in the figure.
}
\label{fig:D4CSL}
\end{figure}

We concentrate on 
the $N_f = 2$ flavor case,
that is the minimum flavor which can couple to the electromagnetic field.
In this case, the $U(1)_{\text{em}}$ electromagnetic field
is introduced in the $\tau^3$ sector of the $U(N_f)_V$ gauge group.
The neutral pion field $\phi$ is also introduced in the $\tau^3$ sector.
Then, in the presence of the background $U(1)_{\text{em}}$
field, the non-zero components of the gauge field in the D8-branes that
correspond to the sine-Gordon kink, are given by
\begin{align}
F_{12} = F_{12}^V 
\frac{\tau^3}{2} = B_3 
\frac{\tau^3}{2},
\qquad
F_{3z} =  f^{-1}_{\pi} 
\frac{\tau^3}{2}
\del_3 \phi \frac{2}{\pi} \, 
 \frac{U^{-1}_{\text{KK}}}{1+\left(\frac{z}{U_{\text{KK}}}\right)^2},
\label{eq:N2_kink_gauge_field}
\end{align}
where $F_{12}^V = B_3$ is the constant background $U(1)_{\text{em}}$ magnetic
field.\footnote{
A D-brane with such a non-Abelian constant magnetic field is called a magnetized D-brane 
\cite{Bachas:1995ik,Berkooz:1996km,Blumenhagen:2000wh,Angelantonj:2000hi,
Cremades:2004wa,Kikuchi:2023awm,Abe:2021uxb}.
A magnetized D-brane also appears in a D-brane configuration for chiral magnets \cite{Amari:2023gqv}, 
which is also a well-known system admitting a CSL ground state.
}
We find that the following RR coupling is induced on the D8-brane\footnote{
Although the RR fields have been rescaled as $C^{(p)} \to \frac{\kappa_{10}^2 \mu_{6-p}}{\pi} C^{(p)}$ in \cite{Sakai:2004cn}, we use the original form in \cite{Myers:1999ps} in this section.
}
\begin{align}
\mu_8 \frac{\lambda_s^2}{2}
\int \text{Tr} [F \wedge F] \wedge C^{(5)} =& \ 
\mu_8 \frac{\lambda_s^2}{2}
\int \! dx^1 dx^2 B_3 
\int^{\infty}_{-\infty} \! dx^3 \,
f_{\pi}^{-1} \del_3 \phi
\notag \\
& \qquad \times 
\int^{\infty}_{-\infty} \! dz \,
\frac{2}{\pi} \frac{U_{\text{KK}}^{-1}}{1 + \left( \frac{z}{U_{\text{KK}}} \right)^2} \,
\varepsilon^{123z a_1 \cdots a_5} \frac{1}{5!} C^{(5)}_{a_1 \cdots a_5}
\notag \\
=& \ 
\pm \mu_4 
\mathbf{B}_3
\varepsilon^{123z a_1 \cdots a_5} \, \frac{1}{5!} C^{(5)}_{a_1 \cdots a_5},
\label{eq:D4_charge_1-soliton}
\end{align}
where only the relevant integrals are presented and 
$\mathbf{B}_3 = \int \! dx^1 dx^2 \frac{B_3}{2\pi}$ is the magnetic flux in the $(x^1,x^2)$-plane and $a_1,\ldots, a_5 = 0,6,7,8,9$.
Here $\lambda_s = 2 \pi \alpha'$, 
$\mu_p = \frac{1}{(2\pi)^p} g_s^{-1} \alpha^{\prime - \frac{p+1}{2}}$ is the D$p$-brane charge and the sign $\pm$ corresponds to the kink $(+)$ or the anti-kink $(-)$, respectively.
If the flux is quantized $\mathbf{B}_3 = n$, this gives the $n$ D4$(06789)$-branes
charge\footnote{
We denote a D$p$-brane extending along the $(x^0, \ldots,
x^{p})$-directions as D$p(012 \cdots p)$.
}. 
Since we are considering the constant magnetic field that does not give a quantized charge, 
the configuraiton \eqref{eq:N2_kink_gauge_field} corresponds to the D4$(06789)$-brane 
localized in the $(x^3,z)$-directions and spread out and
uniformly distributed in the $(x^1,x^2)$-plane (the dissolved D4-branes) \cite{Zwiebach:2004tj}  (table~\ref{tb:N2_branes}).
The CSL is a periodic array of the kink along the $x^3$-direction and it
is therefore identified with the periodic dissolved D4-branes (fig.~\ref{fig:D4CSL}).

We stress that only when both the background magnetic field and the kink are present simultaneously,
they induce the charge of the D4$(06789)$-brane.
In order to see this fact, let us consider the kink without the background magnetic field.
This gives the following RR coupling:
\begin{align}
\mu_8 \lambda_s \int \! \mathrm{Tr} [F] \wedge C^{(7)}
=& \ \mu_8 \lambda_s \int^{\infty}_{-\infty} \! d x^3 \,
f_{\pi}^{-1} \del_3 \phi
\notag \\
& \qquad \times 
\int^{\infty}_{-\infty} \! dz \,
\frac{2}{\pi}
\frac{U_{\text{KK}}^{-1}}{
1 + \left( \frac{U_{\text{KK}}}{z} \right)^2
}
\,
\mathrm{Tr} \Bigg[ \frac{\tau^3}{2} \Bigg]
\,
\varepsilon^{3z a_1 \cdots a_7} \frac{1}{7!} C^{(7)}_{a_1 \cdots a_7}
\notag \\
=& \ 
\mu_6 \,
\mathrm{Tr} 
\left(
\begin{array}{cc}
1 & 0 \\
0 & -1
\end{array}
\right)
\,
\varepsilon^{3z a_1 \cdots a_7} \frac{1}{7!} C^{(7)}_{a_1 \cdots a_7},
\end{align}
where $a_1, \cdots, a_7 = 0,1,2,6,7,8,9$.
The diagonal components of $\tau^3$ 
correspond to the charges of a D6(0126789)-brane and an anti-D6(0126789)-brane, 
which cancel out in the trace, yielding a vanishing net charge.
The pair-canceling charges carried by the kink 
suggest that the excitation consists of a brane/anti-brane pair.
Tachyons from open strings stretched between the brane and the anti-brane suggests that 
an isolated kink in the absence of background magnetic fields is expected to encompass potential instability
and thus can not represent a true ground state.
This picture is consistent with the fact that the $\pi^0$-kink without the background magnetic field is unstable in ChPT \cite{Son:2007ny}.
We also note that the D8-D4 system that results from the kink in the background magnetic field 
is stable since there is no open string tachyon \cite{Polchinski:1998rr}.
This suggests that the kink in the background magnetic field is the stable ground state.

\begin{table}[t]
\begin{center}
\begin{tabular}{c|c|c|c|c|c|c|c|c|c|c}
 & 0 & 1 & 2 & 3 & $y$ & $z$ & 6 & 7 & 8 & 9 \\
\hline
\hline
$N_f$ D8 & $\circ$ & $\circ$ & $\circ$ & $\circ$ &  & $\circ$ & $\circ$ & $\circ$ & $\circ$ & $\circ$ \\
\hline
\hline
(CSL + $B_3$) = D4 & 
$\circ$ & \cellcolor{gray!20} & \cellcolor{gray!20} &  
&
&
& 
$\circ$ 
&
$\circ$ 
&
$\circ$ 
&
$\circ$ 
 \\
\end{tabular}
\end{center}
\caption{The brane configuration corresponding to the CSL in the
 background magnetic field ($N_f = 2$).
The symbol $\circ$ indicates the worldvolume directions of the associated branes.
The shaded region are the dissolved directions.
}
\label{tb:N2_branes}
\end{table}

The sine-Gordon kink in the background field corresponds to the gauge field in 
eq.~\eqref{eq:N2_kink_gauge_field}.
This is localized in the $(x^3,z)$-plane and it is a soliton of codimension two, 
and thus is a vortex membrane in five dimensions.
Note, however, that this does not have 
a conventional vortex charge since
$\text{Tr} [F_{3z}] = 0$ but instead has $\text{Tr} [F_{3z} \tau_3] \neq 0$.
Since the bulk theory is the pure Yang-Mills theory, this vortex should be regarded as a center vortex. 
In a general setup, a center vortex carries a non-Abelian flux of 
$F_{3z}^U  = U F_{3z} U^\dagger$ with 
a particular vortex configuration $F_{3z}$ 
and a gauge transformation $U \in SU(N)$.
However, with the background gauge field $F_{12} = B_3 \tau_3/2$ 
and 
$F_{3z}$ in eq.~(\ref{eq:N2_kink_gauge_field}),
the energy should be minimized for $U={\bf 1}_N$ due to the Chern-Simons term 
in eq.~\eqref{eq:D8CS_term} on the D8-branes.

\begin{figure}[t]
\begin{center}
\includegraphics[scale=.47]{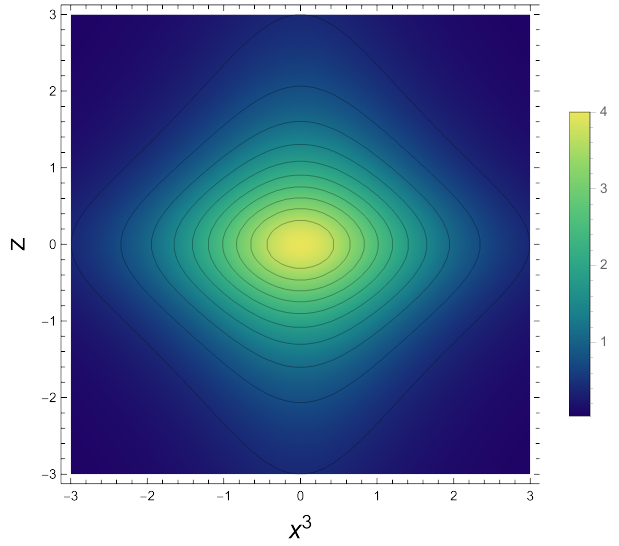}
\includegraphics[scale=.47]{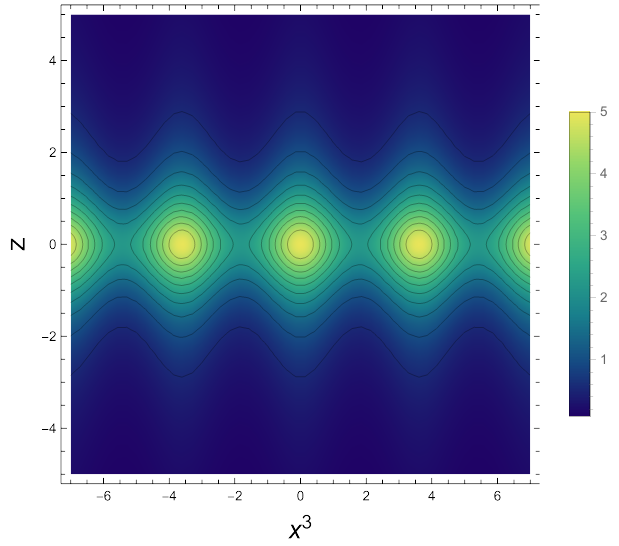}
\includegraphics[scale=.47]{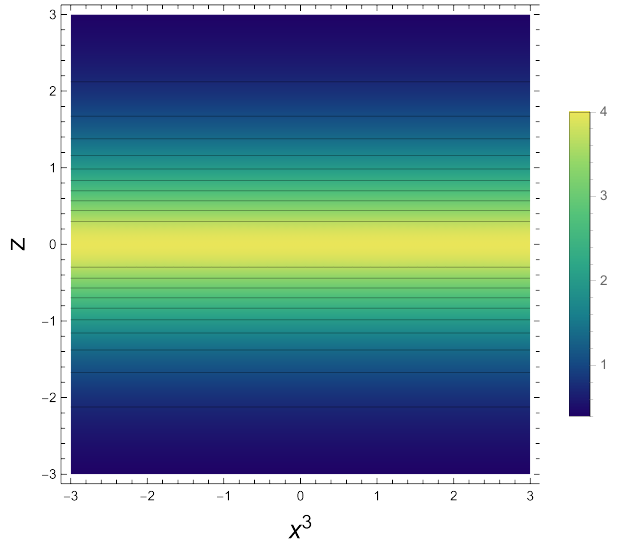}
\end{center}
\caption{
The instanton density $\text{Tr} [F \wedge F]$ in the $(x^3,z)$-plane.
The single soliton (left) and the CSL (middle). 
The parameters are set to $f_{\pi} = B_3 = U_{\text{KK}} = m_\pi = 1$.
The $m_\pi \to 0$ behavior of the kink is also shown (right).
}
\label{fig:CSL_instanton}
\end{figure}

Likewise, it carries 
a non-zero instanton density
$\text{Tr} [F \wedge F]$ (fig.~\ref{fig:CSL_instanton}).
Since the gauge field associated with the CSL in the background field 
does not satisfy the self-duality condition and is not localized in four
dimensions, it is not a conventional instanton. 
Furthermore, it is well known that Skyrmions are identified with instantons in five dimensions \cite{Son:2007ny}.
At the same time, the instantons are identified with the
D4-branes in the holographic model \cite{Sakai:2004cn, Sakai:2005yt}.
Integrating these facts, 
the conventional Skyrmion corresponds to the undissolved D4-brane while
the CSL in the background field is identified with
dissolved periodic D4-branes (See figs.~\ref{fig:D4CSL} and \ref{fig:CSL_instanton}).
Interestingly, both of them has the baryon
charge that appears as the D4-brane charge in the holographic model \cite{Sakai:2004cn, Sakai:2005yt}.
We will comment on this fact in the next subsection.

Before closing this subsection, let us make a comment on a related model.
A comparable system involves vortices being absorbed into a domain wall, which are analogous to Josephson vortices (or fluxons) observed in the Josephson junctions of superconductors in condensed matter physics \cite{Nitta:2012xq,Nitta:2015mma,Nitta:2015mxa,Fujimori:2016tmw}. 
Specifically, a non-Abelian domain wall is permitted in the Higgs phase of a five-dimensional $U(N)$ gauge theory that is coupled with $2N$ complex Higgs fields in the fundamental representation \cite{Shifman:2003uh,Eto:2008dm}. 
The low-energy effective theory localized on this wall is the $U(N)$ chiral Lagrangian \cite{Eto:2008dm,Eto:2005cc}, which resembles our holographic QCD framework.
In that context, the vortices absorbed by the domain wall are described as sine-Gordon solitons on the wall \cite{Nitta:2012xq,Nitta:2015mma}, drawing a parallel to the chiral solitons in our holographic QCD model.
There are, however, crucial differences. In the condensed matter analogue, the bulk gauge theory is in the Higgs phase 
in which vortices are of Abrikosov-type, unlike in our hologpraphic model 
in which the vortices are center vortices living in the confining phase. 
Furthermore, those vortices are trapped inside the domain wall, whereas in our holographic setup, the curved space traps them at the boundary. 

\subsection{Baryon charge}
In this subsection, we discuss the baryon number associated with the CSL.
The gauge field on the D8-branes is decomposed as 
\begin{align}
A = A^a T^a + \frac{1}{\sqrt{2N_f}} \hat{A}
\end{align}
where $A^a T^a$ and $\hat{A}$ are the $SU(N_f)$ and $U(1)$ parts in the $U(N_f)_V$
gauge field.
With this decomposition, the Chern-Simons term for the D8-branes in the background of the 
D4-branes is evaluated as \cite{Hata:2007mb}
\begin{align}
S^{\text{CS}}_{\text{D8}} =& \ 
\ \frac{N_c}{24 \pi^2} \varepsilon^{abcd} 
\int \! d^4 x dz \, 
\Bigg[
\frac{3}{8} \hat{A}_0 \text{Tr} (F_{ab} F_{cd}) - \frac{3}{2}
 \hat{A}_a \text{Tr} (\del_0 A_b F_{cd}) 
\notag \\
& \qquad \qquad 
+ \frac{3}{4} \hat{F}_{ab} \text{Tr} (A_0 F_{cd})
+ \frac{1}{16} \hat{A}_0 \hat{F}_{ab} \hat{F}_{cd}
- \frac{1}{4} \hat{A}_a \hat{F}_{0b} \hat{F}_{cd}
+ \text{(total derivatives)}
\Bigg],
\end{align}
where we have assumed $N_f = 2$ and $a,b,c,d = 1,2,3,z$. 
From this expression, we find that a non-zero instanton density $\varepsilon^{abcd}
\text{Tr} [F_{ab} F_{cd}]$ in the $SU(2)$ sector couples to the baryon chemical potential
$\frac{\hat{A}_0}{\sqrt{2N_f}} = A_{B 0} = \mu_B$.
On the other hand, as we have shown, since the instanton density $\frac{1}{8 \pi^2} \text{Tr} [F \wedge F]$ gives the D4-brane charge density in the D8-branes, they are naturally identified.
Indeed, the $n$ D4-branes, corresponding to the instanton number $n = \frac{1}{8 \pi^2} \int \! \text{Tr} [F \wedge F]$, give Skyrmions of baryon number $n$ \cite{Sakai:2004cn}.

We have evaluated the D4-brane charge density associated with the 1-kink
in the background magnetic field in eq.~\eqref{eq:D4_charge_1-soliton}.
By the rescaling $B_3 \to e B_3$ where $e$ is the electric-magnetic coupling, we have the D4 charge number per unit $(x^1,x^2)$ area as
\begin{align}
\text{(D4 charge)}/A = \frac{e B_3}{2\pi},
\end{align}
where $A = \int dx^1 dx^2$.
This precisely agrees with the baryon number density 
$\frac{N_B}{A} = \frac{e B_3}{2\pi}$ 
carried by the 1-kink in the background magnetic field and baryon chemical potential \cite{Son:2007ny}.

\section{Bulk analysis of chiral soliton lattices}
\label{sec:bulk_CSL}

In the previous section, we focused on zero modes in the $z$-direction as a leading-order approximation of the gauge field.
This section is devoted to the rigorous construction of the minimum holographic state representing CSL \cite{Son:2007ny, Brauner:2016pko}, building upon the framework established in \cite{Kovensky:2023mye}.
Throughout this section, we focus on the $N_f=2$ case.
We use units where $U_{\text{KK}} = 1$ in the subsequent discussion.\footnote{
In order to recover $U_\mathrm{KK}$, 
we need to rescale as $F_{z\mu}\to U_\mathrm{KK}^{-1}F_{z\mu}$,
$z \to U_\mathrm{KK}z$,
$\lambda \to U_\mathrm{KK}^{-1}\lambda$,
$S\to U_\mathrm{KK}^1S$, and
$f_\pi^2\to U_\mathrm{KK}^{-2} f_\pi^2$.
}

We examine the bulk dynamics of the CSL, taking into account contributions from both massive vector mesons and the magnetic field in the Chern-Simons term.
We found that a constant bulk magnetic field satisfies the equations of motion.
On the boundary theory, the magnetic field modifies the pion's decay constant, rendering it a magnetic field-dependent coupling.
This behavior is observed for both the massless and massive pion cases.
Notably, in the massless case, we find a saturation of baryon and energy densities due to the magnetic field's back-reaction.
In the massive case, we observe that the magnetic field back reaction, inhibits the formation of CSLs.

The behavior of the instanton density $\text{Tr}(F\wedge F)$, for large magnetic fields, leads us to interpret it as two D4-branes held apart by the magnetic field.
The larger the magnetic field, the greater the separation between these two branes.
To understand the holographic construction of a CSL state, we must first clarify the imposed boundary conditions.

\subsection{Boundary conditions}

The time component of the gauge field is well known to correspond to the chemical potential of the boundary theory. 
Accordingly, we impose the boundary condition ${A_0(z = \pm \infty) = \mu_B \mathbf{1}}$. 
Furthermore, the limiting values of the bulk field strengths correspond to external fields. 
For the magnetic field, we impose $F_{12} (z = \pm \infty) = B \frac{\tau^3}{2}$. 
Here, $\mu_B$ and $B$ are constants representing the boundary chemical potential and external magnetic field strength, respectively.

To determine the boundary conditions for the remaining gauge fields, following the approach of \cite{Kovensky:2023mye}, we choose a chiral gauge that matches the bulk gauge. 
The bulk gauge choice $A_z = 0$ implies the transformed $U$ takes the form
\begin{equation}
    U' = P\exp\left(-i \int_{-\infty}^\infty dz\, A_z(x^3, z) \right) = \mathbf{1}
\end{equation}
and that $A_3 \rightarrow A_3' = A_3 - \int \partial_3 A_z\, dz$.
On the boundary, we implement this change with a $U(N_f)\times U(N_f)$ gauge transformation, where $U \rightarrow U' = g_+ U g_-^\dagger = \mathbf{1}$.
The choice of $g_\pm$ is not unique; for convenience, we take
\begin{equation}
    g_+ = g^\dagger, \quad g_- = g \implies g^2 = U\,.
\end{equation}
This ensures $U' = g_+ U g_-^{-1} = \mathbf{1}$. 
A benefit of this choice is that it leads to symmetric or antisymmetric boundary conditions for $A$.

At the boundary, the pion field is given by $U = e^{i \tau^3 f_\pi^{-1} \phi(x^3)}$.
To set $U' = 1$, we apply the global phase shift $g_- = g_+^{-1} = g = e^{i \tau^3 f_\pi^{-1} \phi(x^3)/2}$, so that $U' = g_+ U g_-^{-1} = \mathbf{1} \Leftrightarrow U = g^2$.
As previously noted, $g$ induces a transformation in the $x^3$ component of the gauge field.
On the boundary, this results in $A_3 \propto \partial_3\phi \tau^3$ being nonzero, while the other components remain unchanged. 
This is necessary to maintain consistency with our choice of $g$.

Therefore, the boundary conditions at $z \to \pm \infty$ are
\begin{equation}
    \label{eq:boundary_conditions}
    \begin{aligned}
        A_0(\pm\infty) &= \mu_B\,,  \\ 
        A_1(\pm\infty) &= - \frac 12 B \frac{\tau^3}2 x^2\,, \\
        A_2(\pm\infty) &= + \frac 12 B \frac{\tau^3}2 x^1\,, \\
        A_3(\pm\infty) &= \mp f_\pi^{-1}\partial_3 \phi \frac{\tau^3}{2}\,.
    \end{aligned}
\end{equation}
At the level of the equations of motion, we impose that the bulk fields $A_0$ and $A_3$ depend only on $x^3$ and $z$, and they are assumed to take values in $u(1)$ and the Cartan subalgebra of $u(2)$.
The boundary conditions in the $x^3$ direction remain unspecified; they should be chosen to be consistent with both the boundary theory and the bulk equations of motion.

We also employ the ansatz $A_1 = - \frac{1}{2} \mathbf{b} (z) x^2$ and $A_2 = \frac{1}{2} \mathbf{b} (z) x^1$, so that the field strength is $F_{12} = \frac{1}{2} \mathbf{b}(z)$.
Here, $\mathbf{b} = \hat{b}(z) \frac{\mathbf{1}}{2} + b(z) \frac{\tau^3}{2}$ is assumed to have only $\mathbf{1}$ and $\tau^3$ components.

In summary, the gauge field ansatz reads
\begin{equation}
    A = A_0(x^3, z)\, dt + \frac{1}{2} \mathbf b(z) (x^1 dx^2 - x^2 dx^1) + A_3(x^3, z)\, dx^3\,,
\label{eq:ansatz}
\end{equation}
where the gauge field components $A_0$ and $A_3$ generally have nonzero $\tau^3/2$ and $\mathbf{1}/2$ components in $u(2)$.
Nevertheless, $\mathbf{b} = \mathrm{const.}$ solves the equations of motion for variations in $A_1$ and $A_2$, so we set $\hat{b}(z) = \hat{B}$ and $b(z) = B$.
We keep $B$ and $\hat{B}$ generic until the boundary conditions are explicitly imposed.

\subsection{Equations of motion}

Since the pion mass term does not appear directly as a local term in the bulk gauge field, we first focus on the massless limit as a specific case.
With the ansatz \eqref{eq:ansatz}, the equations of motion for the massless case derived from varying $S^\mathbf{DBI}_\mathrm{D8} + S^\mathbf{CS}_\mathrm{D8} 
$ [eqs.~\eqref{eq:DBI} and \eqref{eq:CS}]
 reduce to
\begin{align}
    \partial_3 \left(K F_{z3} \right) &= -\mathcal{B} F_{30}, \label{eq:masslesspion_eomz0} \\
    \partial_z \left(K F_{z3} \right) &= -\mathcal{B} F_{z0}, \label{eq:masslesspion_eom3}\\
    \partial_z \left(K F_{z0} \right) + \partial_3 \left(K^{-1/3} F_{30} \right)  &= -\mathcal{B} F_{z3},\label{eq:masslesspion_eom0}
\end{align}
where $\mathcal{B} := (54\pi/\lambda) \mathbf{b}$ and $K(z) = 1 + z^2$ as defined in 
eq.~\eqref{eq:K}.
For the massive case,
the mass term is known only in four dimensions, and thus we assume 
the mass term appears  
in the right hand side of eq.~(\ref{eq:masslesspion_eomz0}) 
as
\begin{align}
    \partial_3 \left(K F_{z3} \right) &= -\mathcal{B} F_{30}
    - \frac{2}{\pi} m_\pi^2 \sin(f_\pi^{-1} \phi) \frac{\tau^3}{2}.
     \label{eq:masslesspion_eomz} 
   \end{align}  
The bulk mass action to give this mass term by the variation
is proposed in appendix  \ref{app:mass_action_variation}.

Following \cite{Rebhan:2008ur}, we can simplify these equations further through variable substitutions.
Introducing $f_{\mu \nu} = K F_{\mu\nu} = \hat{f}_{\mu\nu}\frac{\mathbf{1}}{2} + f_{\mu\nu}\frac{\tau^3}{2}$, we can solve for $f_{30}$ and $f_{z0}$ in terms of $f_{z3}$, reducing the system to a single equation for $f_{z3}$:\footnote{We momentarily neglect the “pion” decomposition for clarity of notation.}
\begin{equation}
    \label{eq:massless_fz3ode}
    \partial_z \left(K \partial_z f_{z3} \right) + \partial_3 \left(K^{-1/3} \partial_3 f_{z3} + \frac{2}{\pi} m_\pi^2 \sin(f_\pi^{-1} \phi) \frac{\tau^3}{2}\right) = \mathcal{B}^2 K^{-1} f_{z3}.
\end{equation}
It is important to note that the pion mass action, explicitly represented by the term that is proportional to $m_\pi^2$, is included here, despite it coming from a local action.
The reason why it appears here locally is because, when varied the mass action action's content is located at two points along $z$ - the boundaries.
Within eq.~\eqref{eq:massless_fz3ode} the mass and magnetic field are fully incorporated.
The effects of mass will be discussed in section~\ref{sec:bulkmn0}.

It is convenient to decompose $\hat{f}_{z3}$ 
and $f_{z3}$ into their $\tau^\pm = \mathbf{1} \pm \tau^3$ components:
$\hat{f}_{z3} = f_+ + f_-$ and $f_{z3} = f_+ - f_-$.
This allows us to rewrite the equations as two decoupled equations:
\begin{equation}
    \label{eq:massless_fpmode}
    \partial_z (K \partial_z f_{\pm}) + K^{-1/3} \partial_3^2 f_{\pm} - K^{-1} \mathcal{B}_\pm^2 f_{\pm} = \pm \frac{1}{\pi} m_\pi^2 \sin(f_\pi^{-1} \phi),
\end{equation}
where $\mathcal{B}$ is decomposed into its $\tau^{\pm}$ components,
$\mathcal{B}_\pm := \frac{1}{2}\left(\mathcal{B}_0 \pm \mathcal{B}_3\right)$.
Next we construct the on-shell action as to find the ground state.

\subsection{Effective boundary action}

To analyze the boundary dynamics, we can perform a partial integration of the action. This is manageable if we express our fields in a separable form. Although the $\mathbf{b}^2$ term in $S^\mathbf{DBI}_\mathrm{D8}$ diverges, we neglect it since it represents an external field on the boundary.

The DBI action includes contributions from three fields:
\begin{equation}
    S^\mathbf{DBI}_\mathrm{D8} = \frac \pi4 f_\pi^2\int d^4 x d z 
    \text{Tr} \left( K^{-1/3} F_{30}^2 + K (F_{z0}^2 - F_{z3}^2) \right).
\end{equation}
Additionally, there are terms involving $F_{30}$ and $F_{z0}$.
The CS term is similar but does not contain the $F_{30}$ term:

\begin{equation}
    S^\mathbf{CS}_\mathrm{D8} =  
     \frac {N_c}{12\pi^2f_\pi}\int d^4x\mu_B B\del_3\phi + \frac{\pi f_\pi^2}{12} \mathcal B_3 \int d^4 x
\end{equation}
The extra term here is a second ``non-boundary'' contribution.

As previously noted, there is a distinction between the ChPT theory and the bulk-derived theory discussed here.
The difference arises because we have incorporated the effects of the magnetic field into the bulk dynamics.
In the 't Hooft limit, $\lambda \to \infty$, extra terms vanish and the resulting theory matches the one dual to the ChPT. At this point, we can write the effective boundary action:
\begin{equation}
    \label{eq:effective_boundary_action}
    \begin{aligned}
        S_{\partial\mathcal M} &=
        \frac\pi4f_\pi^2 \int d^4x\, dz\, \text{Tr} \left( K^{-1/3} F_{30}^2 + K (F_{z0}^2 - F_{z3}^2) \right) \\ &+ 
        \frac {N_c}{12\pi^2f_\pi}\int d^4x\mu_B B\del_3\phi + \frac{\pi f_\pi^2}{12} \mathcal B_3 \int d^4 x dz \text{Tr}(A_3 F_{z0} \tau^3)\\
    \end{aligned}
\end{equation}
where recall that $\mathcal B _3 = (54\pi/\lambda) B = (N_c \pi^3/f_\pi^2) B$.

\subsection{Case: $\mathcal B_3 \neq 0$ and $m_\pi = 0$}
\label{sec:bulkbn0m0}

Here, we analyze the case of a non-vanishing magnetic field with zero pion mass.
This case is relatively straightforward because the pion mass action does not need to be considered.
Applying the results from appendix~\ref{app:solution_m0Bn0}, the solution to the boundary value problem~\eqref{eq:massless_fz3ode} is
\begin{equation}
\begin{aligned}
    F_{z3} &= -\frac{1}{K} \frac{\mathcal B_3\,\del_3\phi}{2 f_\pi \sinh\left(\frac{\pi}{4}\mathcal B_3\right)} \cosh\left(\frac{1}{2}\mathcal B_3 \arctan{z} \right) \frac{\tau^3}{2} \stackrel{\mathcal B_3 \to 0}{\approx} -\frac{1}{K} \frac{2}{\pi} \frac{\del_3\phi}{f_\pi} \frac{\tau^3}{2},\\
    F_{z0} &= \frac{1}{K} \frac{\mathcal B_3\,\del_3\phi}{2 f_\pi \sinh\left(\frac{\pi}{4}\mathcal B_3\right)} \sinh\left(\frac{1}{2}\mathcal B_3 \arctan{z} \right) \frac{\mathbf 1}{2}, \\
    F_{30} &= 0, \\
    A_3 &= -\frac{\del_3\phi}{f_\pi \sinh\left(\frac{\pi}{4}\mathcal B_3\right)} \sinh\left(\frac{1}{2}\mathcal B_3 \arctan{z} \right) \frac{\tau^3}{2},\\
    A_0 &= \frac{\del_3\phi}{f_\pi \sinh\left(\frac{\pi}{4}\mathcal B_3\right)} \left(\cosh\left(\frac{1}{2}\mathcal B_3 \arctan{z} \right) - \cosh\left(\frac{\pi}{4}\mathcal B_3  \right)\right) \frac{1}{2} + \mu_B \mathbf 1.\\
\end{aligned}
\end{equation}
Let us now evaluate the effective boundary Hamiltonian, eq.~\eqref{eq:effective_boundary_action}.
The result is given by
\begin{equation}
    \label{eq:effective_boundary_action_m0Bn0}
    \begin{aligned}
    H_{\partial \mathcal M} &= 
    \int d^4x \left(
     \frac{1}{2} \frac{\tilde{f}_\pi^2}{f_\pi^2} (\partial_3 \phi)^2 
    - \frac {N_c}{12\pi^2f_\pi}\mu_B B\del_3\phi 
    \right). \\
    \end{aligned}
\end{equation}
Effectively, the pion decay constant acquires a $B$-dependence, denoted by $\tilde{f}_\pi$.
An analytical expression for $\tilde f_\pi$ can be obtained:
\begin{equation}
    \frac{\tilde{f}_\pi^2}{f_\pi^2} = \frac{\pi  \mathcal{B}_3^2 }{16 \sinh^2\left(\frac{\pi  \mathcal{B}_3}{4}\right)} \int dz \frac{1}{K} \left(1+\frac{8}{3} \sinh ^2\left(\frac{1}{2} \mathcal{B}_3 \arctan z \right)\right)
\end{equation}
After evaluating the integrals, the ratio $\tilde{f}_\pi^2 / f_\pi^2$ becomes
\begin{equation}
    \label{eq:effective_decay_constant}
    \frac{\tilde{f}_\pi^2}{f_\pi^2} = \frac{\pi  \mathcal{B}_3}{48  \sinh^2\left(\frac{\pi  \mathcal{B}_3}{4}\right)} \left(8 \sinh \left(\frac{\pi  \mathcal{B}_3}{2}\right)-\pi  \mathcal{B}_3\right)
\end{equation}
where  $f_\pi^2 = 4\kappa/\pi$ in our conventions.
For large $\mathcal B$, we have
\begin{equation}
    \tilde{f}_\pi^2 \sim \frac{1}{3} \pi  \mathcal{B}_3 f_{\pi}^2 = \frac{B N_c}{3 \pi ^2}
\end{equation}
This finding is in agreement with large magnetic fields \cite{Shushpanov:1997sf,Simonov:2015xta} where the effective pion decay constant goes like $\sqrt{B}$ for large magnetic field.
In addition, for small $\mathcal B_3$,
\begin{equation}
    \tilde{f}_\pi^2 \sim f_{\pi }^2 \left(1 + \frac{5}{144} \pi ^2 \mathcal{B}_3^2\right) = f_{\pi }^2 +\frac{5 B^2 N_c^2}{144 \pi^4  f_{\pi }^2}
\end{equation}
This is in agreement with the results from refs.~\cite{Shushpanov:1997sf,Simonov:2015xta}since they find quadratic behavior for weak magnetic fields.

\paragraph{Ground State}

\begin{figure}[t]
    \begin{center}
        \includegraphics[width=0.5\textwidth]{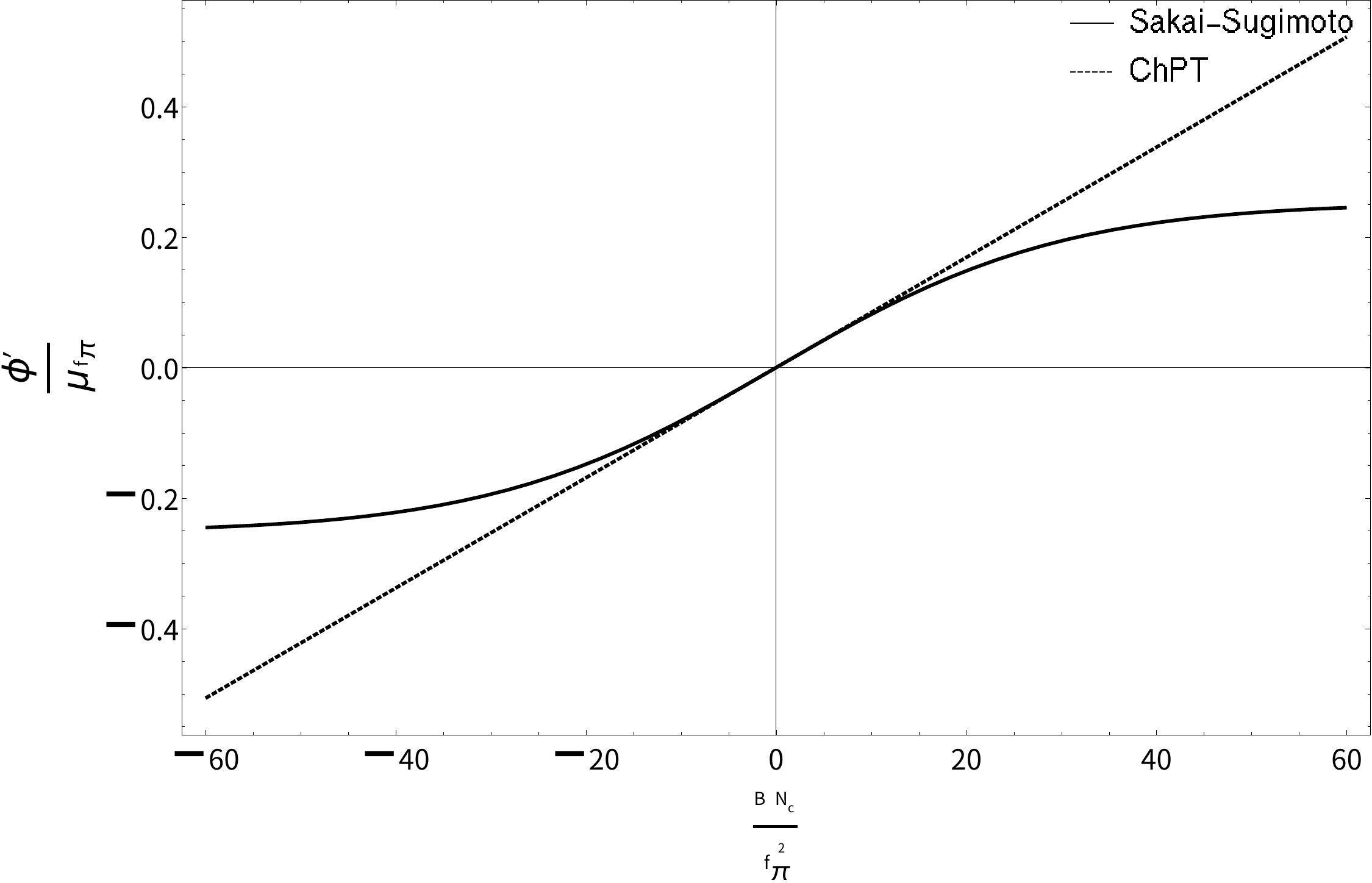}
    \end{center}
    \caption{
    The gradient of the ground state for the massless case.}
    \label{fig:WSS_deloverb}
\end{figure}

For general values of $\mathcal B_3$, we can minimize the energy with respect to $\phi$.
The result is given by
\begin{equation}
    \label{eq:bulkboundary_bn0m0}
    \phi = \frac{\mu_B B N_c}{12\pi^2 f_\pi} \frac{f_\pi^2}{\tilde{f}_\pi^2} \left( x^3 - x^3_0 \right),
\end{equation}
where $x^3_0$ is a constant.
Recall that the ratio $f_{\pi}^2/\tilde{f}_{\pi}^2$ in eq.~\eqref{eq:effective_decay_constant} depends on $\mathcal{B}_3$ (and hence $B$).
Beyond the extreme limits of small and large $\mathcal B_3$, it is worth noting that the slope is an increasing function of $\mathcal B_3 > 0$.
Thus, the maximum slope is approached as $\mathcal B_3 \to \infty$ 
(see fig.~\ref{fig:WSS_deloverb}).

In the large $\mathcal B_3$ limit, the form of the ground state remains the same, but the value of $\tilde f_\pi$ changes.
Explicitly, the ground state reads:
\begin{equation}
    \phi \sim \frac{\mu_B  f_{\pi }}{4} (x^3 - x^3_0)
\end{equation}
In contrast to ChPT, the SS model predicts that, for large $\mathcal B_3$, the slope of $\phi$ saturates, thereby limiting the energy that the $\phi$ field can sustain.

\subsection{Case: $m_\pi \neq 0$}
\label{sec:bulkmn0}

We now consider the general case where both the magnetic field $\mathcal B \neq 0$ and the pion mass $m_\pi \neq 0$ are nonzero.
This involves solving eq.~\eqref{eq:massless_fpmode} in full.
This equation does not yield to analytical solutions easily and requires a full numerical analysis, which we leave for future work.
However, for a sufficiently small $\mathcal B_3$, we want to show that the additional 
boundary term is subleading in $\mathcal B_3$.
We briefly discuss the $F_{30}^2$ term in the effective boundary action.

\paragraph{The $F_{30}^2$ term}
To consistently include all bulk contributions, we would need to incorporate the $F_{30}$ term.
\begin{equation}
\begin{aligned}
    \frac{\pi}{4}f_\pi^2\int \text{Tr}\, K^{-1/3} F_{30}^2 &= \int d^4x\, \frac{g_\pi^2}{f_\pi^2} \frac{1}{2}(\phi'')^2 \\
    \frac{g_\pi^2}{f_\pi^2} &:= \int dz K^{-1/3}\frac{\pi}{4\sinh^2\left(\frac{\pi}{4}\mathcal B_3 \right)}  \left(\cosh\left(\frac{\mathcal B_3}{2} \arctan{z} \right) - \cosh\left(\frac{\pi}{4}\mathcal B_3 \right)\right)^2 \\
\end{aligned}
\end{equation}
This term is proportional to $(\phi'')^2$ in contrast to the $(\phi')^2$ proportional terms.
For small $\mathcal B_3$, $g_\pi^2 \approx  \beta \frac\pi4 \mathcal B_3^2$ where $\beta \approx 0.386370$.
In this case we can neglect the $g_\pi^2$ term since $g_\pi^2 / \tilde{f}_\pi^2 \ll 1$.
For the ground state, we approximate it by starting with ChPT action \eqref{eq:gravity_pion} but 
with a different kinetic energy coefficient, replacing it as $\frac 12 \to \frac 12\tilde{f}_\pi^2/f_\pi^2$ where 
$\tilde{f}_\pi^2/f_\pi^2$ comes from eq.~\eqref{eq:effective_decay_constant}
from the massless pion, non-vanishing magnetic field case in section \ref{sec:bulkbn0m0}.

\paragraph{Ground State for small $\mathcal B_3$}

The equation of motion becomes
\begin{equation}
      \partial_3^2 \phi = \frac{f_\pi^3}{\tilde f_\pi^2 }  m_\pi^2 \sin( f_\pi^{-1} \phi ).
\end{equation}
A solution to this equation can be constructed using Jacobi's amplitude function, yielding a family of solutions as
\begin{align}
    \label{eq:holographic_CSL}
    \phi (x^3) = f_\pi \left[ \pi \pm 2\,\mathrm{am} \left(\frac{m_\pi}{k} \frac{f_\pi}{\tilde{f}_\pi} x^3,\, k^2\right) \right].
\end{align}
This is the CSL at the boundary.

Effectively, the pion mass is rescaled by the dimensionless parameter $f_\pi/\tilde{f}_\pi$. The effective mass reads
\begin{equation}
    \label{eq:bulk_effectivemass}
    \tilde m_\pi = \frac{f_\pi}{\tilde{f}_\pi} m_\pi.
\end{equation}
To determine the energy minimization condition we can do a factor trick.
We can factor out $f_\pi^2/\tilde{f}_\pi^2$ from the Hamiltonian, so both mass and chemical potential terms effectively become $\mathcal B_3$-dependent where.
Thus, the results from the ChPT Hamiltonian apply, but now with $\mathcal B_3$-dependent effective mass, eq.~\eqref{eq:bulk_effectivemass}, and chemical potential,
\begin{equation}
    \tilde \mu_B = \frac{f_\pi^2}{\tilde{f}_\pi^2} \mu_B\,.
\end{equation}
Therefore, the CSL state is energetically favorable for
\begin{equation}
    \label{eq:bulkCSLexistscond}
    \begin{aligned}
    N_c B\tilde\mu_B > 48\pi \tilde m_\pi f_\pi^2 &\implies \frac{f_\pi}{\tilde{f}_\pi} N_c B \mu_B > 48\pi m_\pi f_\pi^2\\
    &\implies \sqrt{1- \frac{5}{f_\pi^4} \left(\frac{B N_c}{12 \pi^2}\right)^2} N_c B \mu_B \gtrsim 48\pi m_\pi f_\pi^2 \,.\\
    \end{aligned}
\end{equation}
Because $f_\pi/\tilde{f}_\pi < 1$ for small $\mathcal B_3$, the required magnetic field for a CSL to form is larger than the CSL of ChPT.

\paragraph{Case: $\mathcal B = 0$ and $m_\pi \neq 0$}
\label{sec:bulkb0mn0}

Here we consider the case of vanishing magnetic field and non-vanishing pion mass.
In this case we can use the equation of motion for the fields strengths but they are decoupled.
After some analysis in appendix \ref{app:solution_mn0B0}, we find that the only non-vanishing field strength is $F_{z3}$:
\begin{equation}
\begin{aligned}
    F_{z3} &= -\frac{1}{K} \frac{2}{\pi} \frac{\partial_3\phi}{f_\pi} \frac{\tau^3}{2}, \\
    A_3 &= -\frac{2}{\pi} \frac{\partial_3\phi}{f_\pi} \arctan{z} \frac{\tau^3}{2},\\
    A_0 &= \mu_B \mathbf 1.
\end{aligned}
\end{equation}
This solution is consistent with the results of section~\ref{sec:chiral_model} and agrees exactly with ChPT.

\subsection{Instanton density}

In this section we examine the instanton density, $\text{Tr}\left(F\wedge F\right)$.
We compare with the previous result in section~\ref{sec:brane_CSL}.

Our component $F \wedge F$ can be written as follows:
\begin{equation}
    \begin{aligned}
        \text{Tr}(F\wedge F) = \frac{B_3}{8} \text{Tr} \big( 
        &\tau^3 F_{3z} dx^1\wedge dx^2\wedge dx^3\wedge dz \\+ 
        &\tau^3 F_{0z} dx^1\wedge dx^2\wedge dx^0\wedge dz \\+ 
        &\tau^3 F_{03} dx^1\wedge dx^2\wedge dx^0\wedge dx^3  \big).
    \end{aligned}
\end{equation}
The trace with $\tau^3$ causes all but the $123z$ component to vanish:
\begin{equation}
    \label{eq:bulkFFcomponent}
    \left(\text{Tr}(F\wedge F)\right)_{123z} = \frac{1}{K}\frac{\mathcal B_3\del_3\phi}{2 f_\pi\sinh\left(\frac\pi4\mathcal B_3\right)}\cosh\left(\frac12\mathcal B_3 \arctan{z} \right).
\end{equation}

\begin{figure}[th]
\begin{center}
\includegraphics[width=0.5\textwidth]{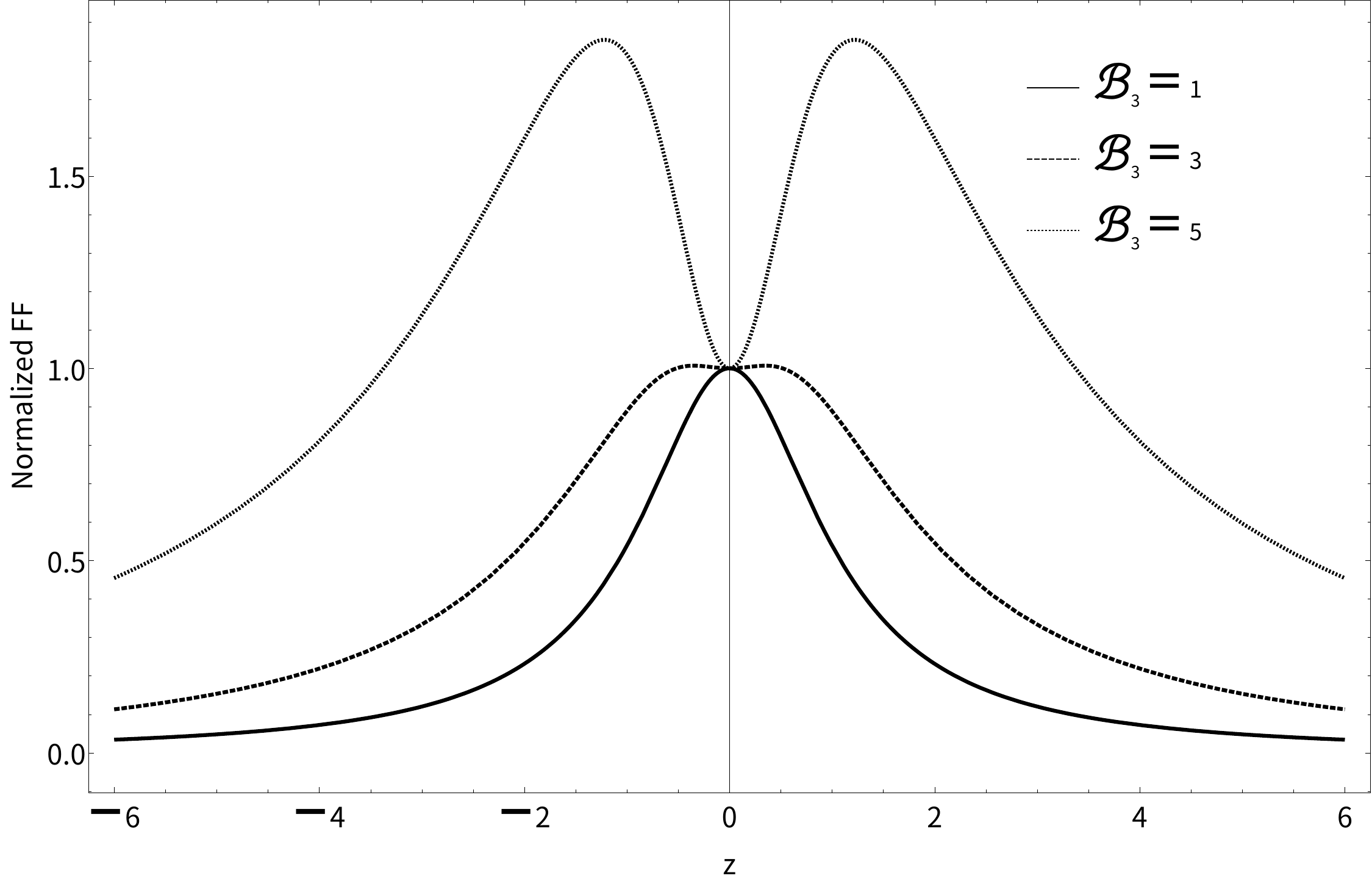}
\end{center}
\caption{
Normalized cross sections of $\text{Tr}\left(F\wedge F\right)$ along the $z$-direction.
The quantities are normalized such that at $z=0$, $\text{Tr}\left(F\wedge F\right) = 0$.
For sufficiently large magnetic fields, one can see the onset of two peaks.
The peaks grow in size relative to the value at the origin.
If the normalization were removed, the peaks would be larger because of the hyperbolic scaling of $F_{z3}$.
}
\label{fig:WSS_ZProfile}
\end{figure}

To better understand the brane embedding, we can focus on the $z$-dependent cross sectional behavior of eq.~\eqref{eq:bulkFFcomponent}.
There are two factors here. 
\begin{equation}
    K^{-1} = (1+z^2)^{-1}\quad\text{and}\quad\cosh\left(\frac12\mathcal B_3 \arctan{z} \right)
\end{equation}
Since these two factors are positive, along with the positive proportionality constant, eq.~\eqref{eq:bulkFFcomponent} is a negative quantity.
Two local extrema of eq.~\eqref{eq:bulkFFcomponent} exist when $\mathcal B_3 > \sqrt{8}$.
Due to the parity of eq.~\eqref{eq:bulkFFcomponent}, these two extrema are located symmetrically
(fig.~\ref{fig:WSS_ZProfile}).

The presence of these minima suggests that for strong enough magnetic fields, two embedding D4-branes can be held apart along the $z$ direction, against gravity.

\begin{figure}[t]
\begin{center}
    \includegraphics[width=.45\textwidth]{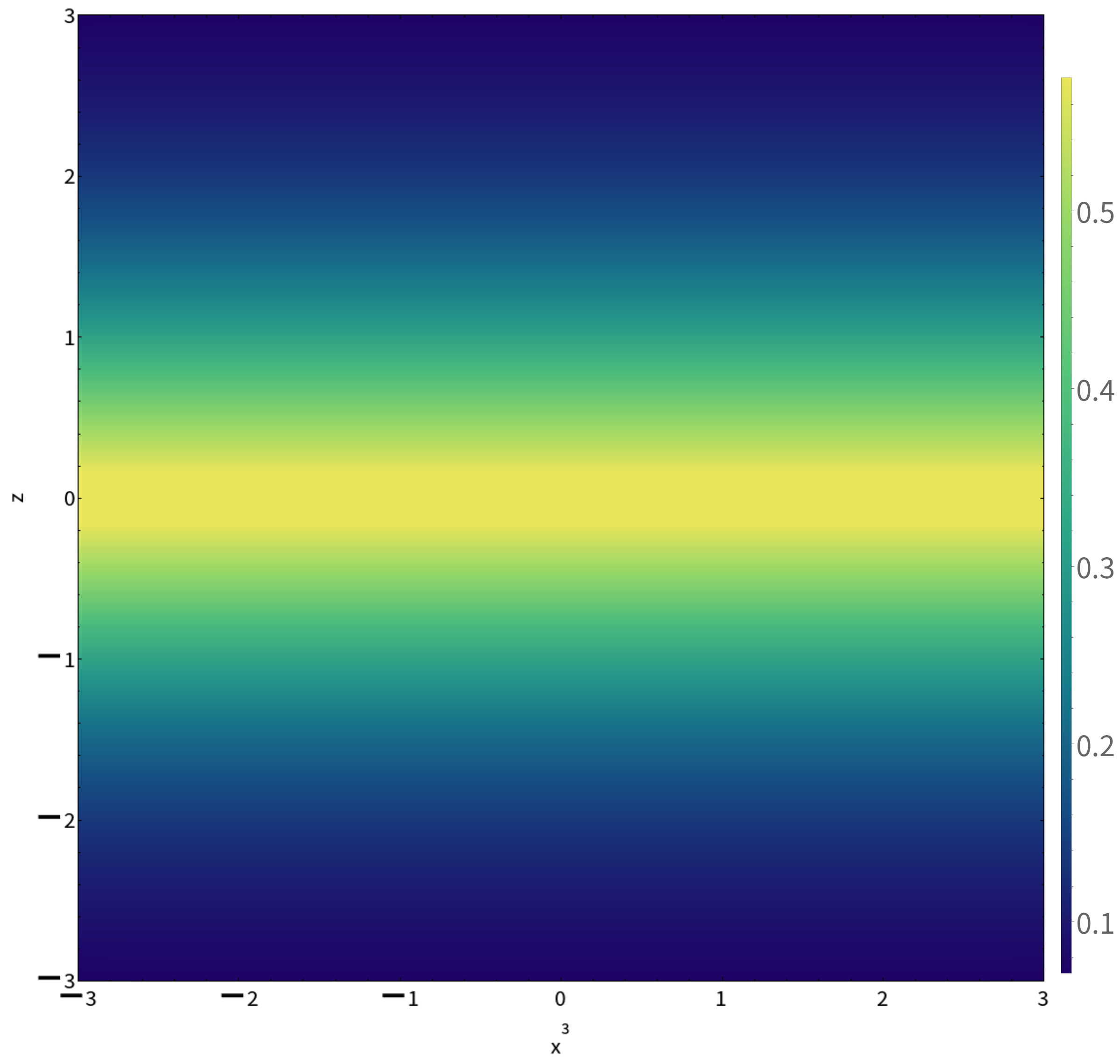}
    \includegraphics[width=.45\textwidth]{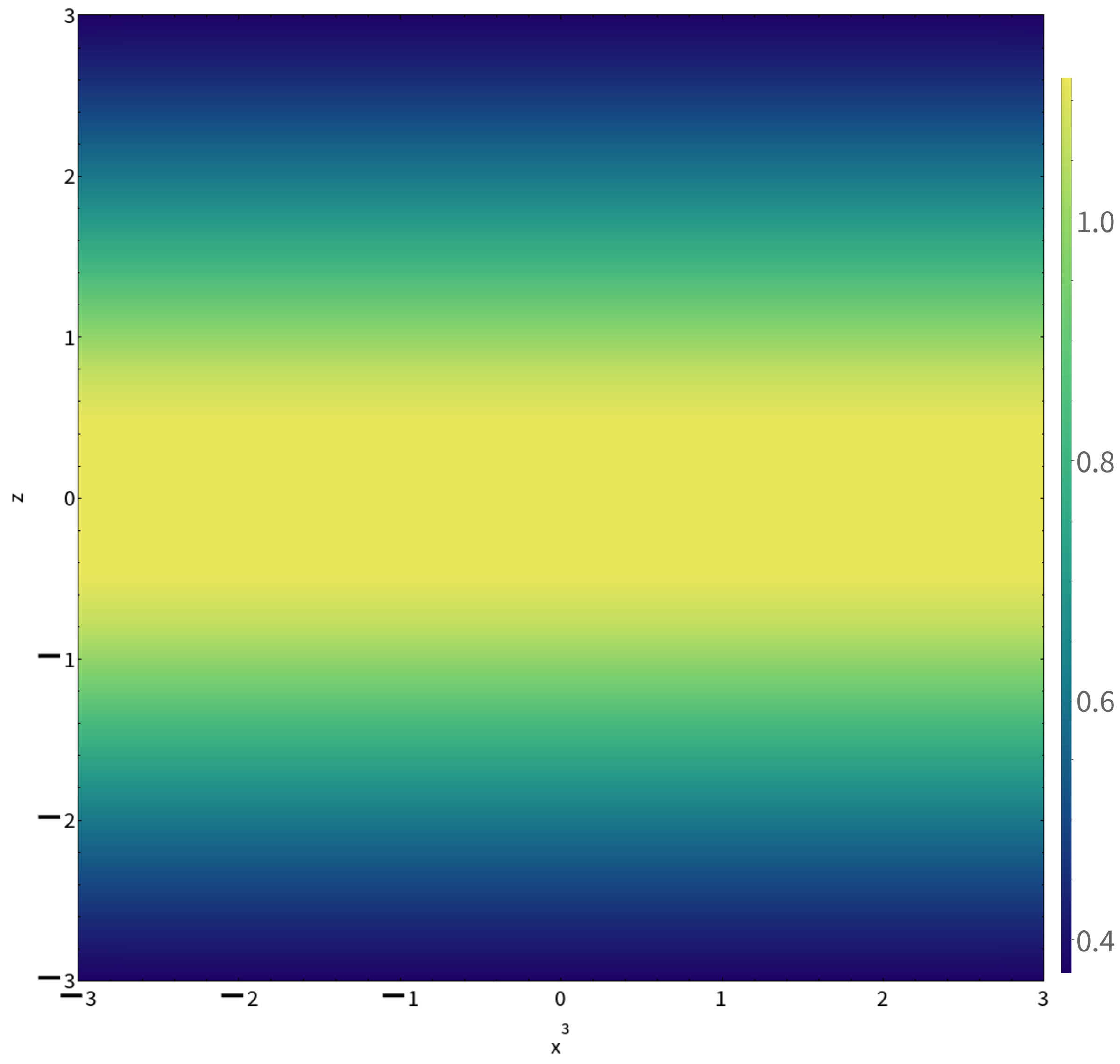}
\end{center}
\caption{
The instanton density magnitude, $\left|\text{Tr} [F \wedge F]\right|$, in the $(x^3,z)$-plane with back reaction for two $m_\pi=0$ kink configurations.
Plots are made with $\mu_B = f_\pi = 1$.
Left plot: $\mathcal B_3 = 1$; right plot: $\mathcal B_3 = 3$.
}
\label{fig:WS_kink}
\end{figure}

\begin{figure}[th]
\begin{center}
    \includegraphics[width=.45\textwidth]{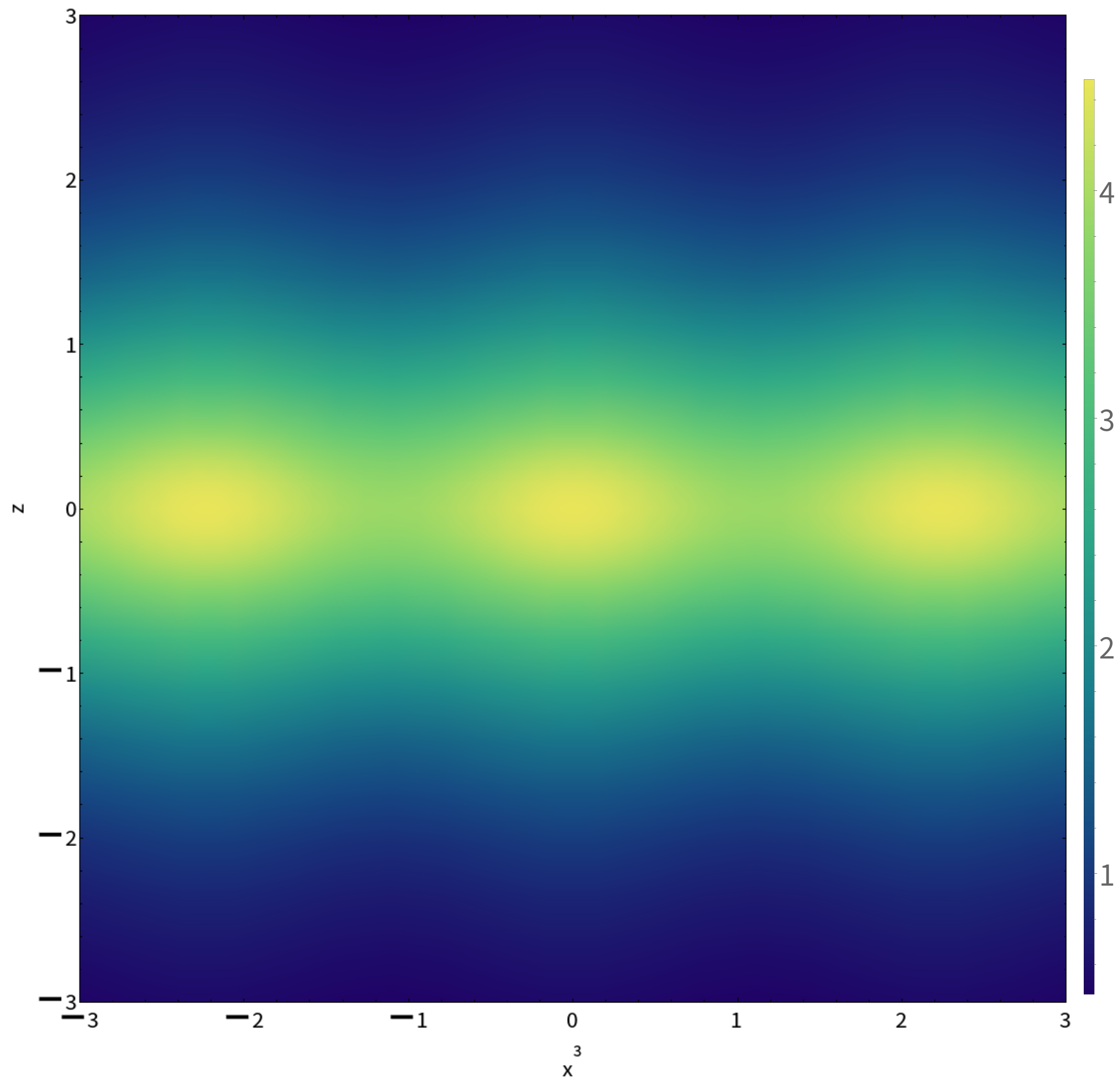}
    \includegraphics[width=.45\textwidth]{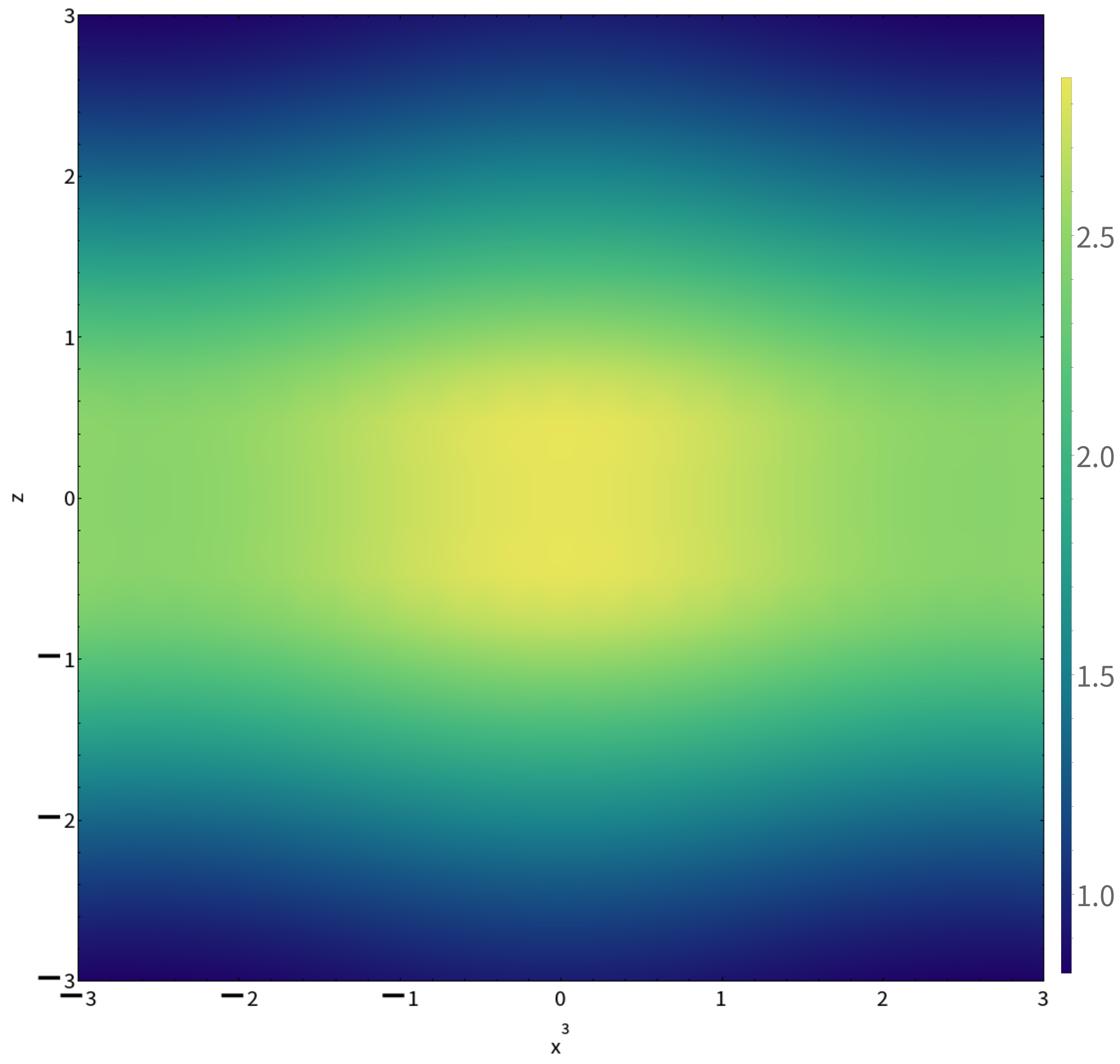}
\end{center}
\caption{
The instanton density magnitude, $\left|\text{Tr} [F \wedge F]\right|$, in the $(x^3,z)$-plane with back reaction for two CSL configurations.
The pion mass is set to $m_\pi = 1$ and the elliptic modulus is set to $k = 1/2$ for illustration.
Plots are made with $\mu_B = f_\pi = 1$.
Left plot: $\mathcal B_3 = 1$; right plot: $\mathcal B_3 = 3$.
}
\label{fig:SS_CSL}
\end{figure}

Comparing with fig.~\ref{fig:CSL_instanton}, we can understand the effects of the CS term on the bulk state.
The key difference between the kink configurations shown in fig.~\ref{fig:WS_kink} and the CSL configurations in fig.~\ref{fig:SS_CSL} lies in the spatial distribution of the instanton density.

In the kink case with $m_\pi = 0$, the instanton density exhibits a more localized structure along the $x^3$ direction, with sharp transitions characteristic of domain wall solutions.
By contrast, the CSL configurations with finite pion mass $m_\pi = 1$ display a more extended, modulated pattern that reflects the underlying chiral spiral structure. 
The CS term introduces additional topological contributions that modify the field strength tensor, leading to these distinct signatures in the instanton density. 
As the magnetic field strength increases from $\mathcal B_3 = 1$ to $\mathcal B_3 = 3$, both configurations show enhanced localization in the $z$ direction, consistent with the emergence of the double-minimum structure predicted by our analytical result in eq.~\eqref{eq:bulkFFcomponent}. 
This behavior demonstrates that the CS term not only influences the ground state configuration but also determines the response of the system to external magnetic fields, providing a direct probe of the underlying topological structure of the bulk theory.

\section{Conclusion and discussions}
\label{sec:conclusion}

In this paper, we have studied the CSL in the
framework of holographic QCD.
By introducing the quark mass deformation and the boundary
conditions corresponding to the strong background magnetic field and
the baryon number density,
we have shown that the CSL becomes a ground state in the gravity dual of QCD
under the appropriate setup.
We have considered the brane configurations corresponding to the CSL. 
The meson field configuration is interpreted as a non-trivial excitation
of the gauge field on the D8-branes. 
The chiral soliton of the meson field looks like a vortex membrane in 
five-dimensional bulk, 
and the CSL is therefore a periodic vortex membrane. By introducing a
constant background field, the vortex becomes instanton-like, 
as $\text{Tr}[F \wedge F]$ is non-zero and the CSL in the background
becomes a source of the D4-branes.
Since the constant magnetic field itself represents a dissolved brane, we
have shown that the CSL in the background magnetic field behaves like a
dissolved D4-brane.
This is compared with the fact that the Skyrmion in the chiral theory
behaves like an instanton in five dimensions and is interpreted as a
localized D4-brane.
Skyrmions are identified with baryons and the D4-brane charge is
interpreted as a baryon charge. 
Indeed, we have shown that by observing the D4 charges, we correctly
reproduced the baryon number density carried by the CSL.
We have found that chiral solitons correspond to  non-self-dual instanton vortices in five dimensions.
These objects exist in pure Yang-Mills theory and thus should be identified with center vortices. 

In addition to our analysis of the brane realization and topological structure, we performed a comprehensive bulk study by solving the full five-dimensional equations of motion for the gauge fields, accounting for the effects of both magnetic field and pion mass. We systematically derived the effective boundary action by integrating out bulk degrees of freedom, revealing how the magnetic field dynamically modifies the effective pion decay constant, $\tilde{f}_\pi$. This leads to the saturation of baryon and energy densities at large magnetic fields for both massless and massive pions; notably, in the massive case, the backreaction inhibits CSL formation because $\tilde{f}_\pi > f_\pi$ for $B\neq 0$, thus raising the threshold for the energic favorability of the CSL phase compared to standard field theory. Analytic insight is possible for small mass or field, but in general, the inclusion of the mass term complicates the equations for large magentic field and requires numerical treatment.

Through the calculation of the instanton density $\text{Tr}[F \wedge F]$, we established that, at strong magnetic field, the bulk configuration corresponds to two D4-branes held apart, with their separation growing as the field increases. This provides an intuitive bulk picture for the emergent topological structures, and underscores the importance of carefully-formulated $z$-direction boundary conditions, which determine the allowed physical solutions and phases in the boundary theory. In particular, the SS model both recovers known ChPT results in the small field and massless limits, and predicts new phenomena in strong field environments. Notably, the slope of the chiral soliton saturates at large magnetic field, a striking departure from the behavior found in the standard ChPT~\cite{Simonov:2015xta, Shushpanov:1997sf}.

Although our explicit focus has been on pionic CSLs, the bulk holographic formalism is sufficiently general to incorporate both massive vector mesons and, with further development, charged pion effects. This opens the way to systematic study of inhomogeneous (solitonic) ground states and the broader QCD phase structure in strong background fields. Future work will address these extensions in more detail.

While we have discussed only the CSL in this paper, it is known, as mentioned in the introduction, that the CSL transitions to the domain-wall Skyrmion phase at higher densities and/or stronger magnetic fields.
Domain-wall Skyrmions are composite states consisting of Skyrmions absorbed into a domain wall, first introduced in field theories in 2+1 dimensions~\cite{Nitta:2012xq,Kobayashi:2013ju,Jennings:2013aea} and later in 3+1 dimensions~\cite{Nitta:2012wi,Nitta:2012rq,Gudnason:2014nba,Gudnason:2014hsa,Eto:2015uqa,Nitta:2022ahj}.
Skyrmions in the bulk are absorbed into a chiral soliton to form topological lumps (or baby Skyrmions), which are supported by $\pi_2(S^2) \simeq \mathbb{Z}$ in the $O(3)$ nonlinear sigma model aapearing on the chiral soliton. 
A single lump within the soliton corresponds to two Skyrmions in the bulk and is therefore a boson~\cite{Amari:2024mip}. 
In holographic QCD, Skyrmions on the boundary correspond to Yang-Mills instanton particles in the five-dimensional bulk~\cite{Hata:2007mb}, as indicated by the Atiyah-Manton description of instantons~\cite{Atiyah:1989dq,Eto:2005cc}.
We therefore expect that Yang-Mills instanton particles in the five-dimensional bulk can be trapped by vortex membranes, forming vortex-instanton composites in which both the vortex membranes and the instantons carry instanton charge.\footnote{
Similar vortex-instanton composites are known to appear when the bulk gauge theory is in the Higgs phase~\cite{Eto:2004rz,Fujimori:2008ee}.
Thus, our vortex-instantons are pure Yang-Mills analogues of these objects.
See also the comment in the last paragraph of section.~\ref{sec:CSL-brane}.
}
These composites manifest as domain-wall Skyrmions at the four-dimensional boundary, where both the chiral solitons and Skyrmions carry baryon number.
Exploring the domain-wall Skyrmion phase in holographic QCD is one of the important directions for future research.

\begin{acknowledgments}
We would like to thank Yuki Amari for the discussion at the early stages of this research project.
We would also like to thank Shigeki Sugimoto and Anton Rebhan for helpful advices and remarks.
This work is supported in part by 
 Japan Society for the Promotion of Science (JSPS) KAKENHI [Grants No. JP22H01221 and JP23K22492 (ME and MN), JP25K07324 (SS)] 
 and the WPI program ``Sustainability with Knotted Chiral Meta Matter (WPI-SKCM$^2$)'' at Hiroshima University (ME and MN).
 MA was provided funding by the National Institute of Technology, Oyama College and initially supported by the JSPS Postdoctoral Fellowship for Research in Japan (Short-term). 
\end{acknowledgments}

\appendix

\section{Calculation of the WZW term} \label{sec:Z_calculation}
In this appendix, we derive the expression of the WZW term appearing in the literature 
\cite{Son:2007ny} from the term $Z$ in  eq.~\eqref{eq:D8CS_term}.
We set
\begin{align}
&
U = \Sigma,
\qquad
U^{-1} = \Sigma^{\dagger},
\qquad
L = \Sigma d \Sigma^{\dagger} = - d \Sigma \, \Sigma^{\dagger},
\qquad
R = d \Sigma^{\dagger} \, \Sigma = - \Sigma^{\dagger} \, d \Sigma,
\notag \\
&
A_L = A_R = A = A_Q Q + \tilde{A}_B \mathbf{1},
\qquad
d A = Q d A_Q 
+ d \tilde{A}_B \mathbf{1},
\label{eq:WZW_setup}
\end{align}
where $Q$ is in the Cartan subalgebra of the $U(N_f)_V$, $A_Q$ is the Abelian gauge field component
and $\tilde{A}_B = A_B N_c^{-1}$ is proportional to $\mathbf{1}$.

A careful calculation reveals that 
\begin{align}
Z=& \
- 2 i
A_Q d A_Q \, \mathrm{Tr} 
\Big[
Q^2 (R + L) 
\Big]
- i
A_Q d A_Q \, \mathrm{Tr} 
\Big[
Q d \Sigma^{\dagger} Q \Sigma
-
Q d \Sigma Q \Sigma^{\dagger}
\Big]
\notag \\
& \ 
+
A_Q \, \mathrm{Tr}
\Big[
Q (R^3 + L^3)
\Big]
+
2 \tilde{A}_B \, \mathrm{Tr} 
\Big[
L^3
\Big]
- 6 i
\tilde{A}_B d A_Q \, \mathrm{Tr} 
\Big[
Q (R + L)
\Big]
\notag \\
& \
+ 6 i 
\tilde{A}_B A_Q \, \mathrm{Tr}
\Big[
Q (R^2 - L^2)
\Big]
+
3 i 
\tilde{A}_B d \tilde{A}_B \mathrm{Tr} \Big[ R + L \Big].
\end{align}
%
Using the following relation,
\begin{align}
\tilde{A}_B d \, 
\Big\{
A_Q \, \mathrm{Tr}
\Big[
Q (R + L)
\Big]
\Big\}
= \tilde{A}_B d A_Q \, \mathrm{Tr} 
\Big[
Q (R + L)
\Big]
- \tilde{A}_B A_Q \, \mathrm{Tr} 
\Big[
Q (R^2 - L^2)
\Big],
\end{align}
we obtain the following result:
\begin{align}
 \frac{N_c}{48 \pi^2} \int \! Z =& \ 
\frac{N_c}{48 \pi^2} 
\varepsilon^{\mu \nu \rho \sigma} 
\int \! d^4 x \,
\Bigg\{
A_{Q \mu} \mathrm{Tr} 
\Big[
Q (L_{\nu} L_{\rho} L_{\sigma} + R_{\nu} R_{\rho} R_{\sigma} )
\Big]
+
3 i
\tilde{A}_{B \mu} \del_{\nu} \tilde{A}_{B \rho} \,
\mathrm{Tr} \Big[ R_{\sigma} + L_{\sigma} \Big]
\notag \\
& \qquad \qquad \qquad 
- i F_{Q \mu \nu} A_{Q \rho} \mathrm{Tr}
\Big[
Q^2 (L_{\sigma} + R_{\sigma})
+ \frac{1}{2} Q \Sigma Q \del_{\sigma} \Sigma^{\dagger}
- \frac{1}{2} Q \Sigma^{\dagger} Q \del_{\sigma} \Sigma
\Big]
\Bigg\}
\notag \\
& \ 
- \int \! d^4 x \,
\tilde{A}_{B \mu} 
\,
\left(
-
\frac{1}{24 \pi^2}
\right)
\varepsilon^{\mu \nu \rho \sigma}
\Bigg\{
\mathrm{Tr} 
\Big[
L_{\nu} L_{\rho} L_{\sigma}
\Big]
- 
3 i \,
\del_{\nu}
\Big(
A_{Q \rho} \mathrm{Tr} 
\Big[
Q (R_{\sigma} + L_{\sigma})
\Big]
\Big)
\Bigg\}.
\label{eq:WZW_SS}
\end{align}
If we focus on the $SU(N_f)$ part of $L$ and $R$, the term including $\tilde{A}_{B}$
in the first line of the above equation vanishes.
In this case and when we consider $N_c = 3$, 
\eqref{eq:WZW_SS} reduces to the WZW term elucidated in the literature \cite{Son:2007ny}.

\section{Solving for bulk fields}
\label{app:solution_m0Bn0}
\subsection{Case: $\mathcal B_3\neq0$ and $m_\pi=0$}

In this appendix we will focus on solving eq.~\eqref{eq:massless_fz3ode} and by implication \eqref{eq:massless_fz3ode} for the massless pion case.
First note we can use, a separation of variables ansatz, $f_{\pm} = X_\pm(x^3) Z_\pm(z)$, leads to the following equations:
\begin{equation}
    \label{eq:zode}
    \begin{aligned}
        \left(K Z'_\pm\right)' - k_{\pm}^2 K^{-1/3} Z_\pm - K^{-1} \mathcal{B}_\pm^2 Z_\pm &= 0, \\
        X_\pm'' + k^2_\pm X_\pm &= 0, \\
    \end{aligned}
\end{equation}
where $k^2_\pm$ is the separation constant.
Clearly, $X_{\pm}$ is a linear combination of $\sin (k_{\pm} x^3)$ and $\cos (k_{\pm} x^3)$.

Now observe that eq.~\eqref{eq:masslesspion_eom3} implies that 
\begin{equation}
    \label{eq:fz3proptoA0}
    f_{z3} = -\mathcal B (A_0 - \alpha_0)\,,
\end{equation}
where $\alpha_0$ is a function of $x^3$.
Due to the boundary conditions imposed upon $A_0$, eq.~\eqref{eq:fz3proptoA0} implies $f_{z3}$, must vanish if it has $x_3$ dependence.
This implies that all the $k_\pm$ solutions of eq.~\eqref{eq:zode} must vanish.
This can be seen since the $Z_\pm$ equations of motion, are equivalent to a particle that has a repulsive force act upon it directed away from the origin, $Z=0$.
Therefore the only vanishing $k_\pm \neq 0$ solution is the trivial solution since any other solution cannot have $f_{z3}$ vanish on the boundary.
This allows us to simplify eq.~\eqref{eq:zode} to eq.~\eqref{eq:zode_masslesspion_magneticfield}:
\begin{equation}
    \label{eq:zode_masslesspion_magneticfield}
    \del_z (K \del_z Z_\pm) - K^{-1} \mathcal B_\pm^2 Z_\pm = 0 .
\end{equation}
Since $A_3$ is a $z$ odd function, in general even in $Z_\pm$ must be odd. 
The first step to solve eq.~\eqref{eq:zode_masslesspion_magneticfield}, to introduce the coordinate $\xi = \arctan{z}$.
Also $K\del_z = \del_\xi$.
In this coordinate, eq.~\eqref{eq:zode_masslesspion_magneticfield} can be written as eq.~\eqref{eq:xiode_masslesspion_magneticfield}:
\begin{equation}
    \label{eq:xiode_masslesspion_magneticfield}
    K^{-1}\left(\del_\xi^2 Z_\pm - \mathcal B_\pm^2 Z_\pm\right) = 0. 
\end{equation}

The general even solution to eq.~\eqref{eq:xiode_masslesspion_magneticfield} can be written as
\begin{equation}
    \begin{aligned}
        f_\pm &= \xi_\pm\cosh\left(\mathcal B_\pm \arctan{z} \right)\\
        f_{z3}&= (f_+ + f_-) \frac 12 + (f_+ - f_-) \frac {\tau^3}2  \\
              &= \left((\xi_+ + \xi_-) \frac 12 + (\xi_+ - \xi_-) \frac {\tau^3}2\right) \cosh\left(\frac{\mathcal B_3}{2} \arctan{z} \right) \\
    \end{aligned}
\end{equation}
where we specialized the magnetic field to $\mathbf B = B \frac {\tau^3}2$.
We can immediately integrate to find the $A_3$:
\begin{equation}
    \begin{aligned}
        F_{z3} &= K^{-1}\left((\xi_+ + \xi_-) \frac 12 + (\xi_+ - \xi_-) \frac {\tau^3}2\right) \cosh\left(\frac{\mathcal B_3}{2} \arctan{z} \right) \\
        A_3 &= \frac{2}{\mathcal B_3}\left((\xi_+ + \xi_-) \frac 12 + (\xi_+ - \xi_-) \frac {\tau^3}2\right) \sinh\left(\frac{\mathcal B_3}{2} \arctan{z} \right) + \alpha_3 . \\
    \end{aligned}
\end{equation}
Due to boundary conditions $\alpha = 0$ and 
\begin{equation}
    \xi_+ = -\xi_- = -\frac{\mathcal B_3}{4 \sinh\left(\frac{\pi}{4}\mathcal B_3 \right)} \frac{\del_3\phi}{f_\pi}\,.
\end{equation}
Now we write $A_3$ in it's final bulk form
\begin{equation}
    \begin{aligned}
        F_{z3} &= -K^{-1} \frac{\mathcal B_3 \del_3\phi}{2 f_\pi\sinh\left(\frac{\pi}{4}\mathcal B_3 \right)}   \cosh\left(\frac{\mathcal B_3}{2} \arctan{z} \right) \frac{\tau^3}2 \\
        A_3 &= -\frac{\del_3\phi}{f_\pi \sinh\left(\frac{\pi}{4}\mathcal B_3 \right)}   \sinh\left(\frac{\mathcal B_3}{2} \arctan{z} \right) \frac{\tau^3}2 .\\
    \end{aligned}
\end{equation}
Using eq.~\eqref{eq:fz3proptoA0}, we can solve for $A_0$:
\begin{equation}
        A_0 = \frac{\del_3\phi}{f_\pi \sinh\left(\frac{\pi}{4}\mathcal B_3 \right)}   \cosh\left(\frac{\mathcal B_3}{2} \arctan{z} \right) \frac 12 + \alpha_0
\end{equation}
where $-\mathcal B^{-1} = -2\mathcal B_3^{-1}\tau^3$.
Nevertheless, due to boundary conditions 
\begin{equation}
\begin{aligned}
        \alpha &= \left(2\mu_B +\frac{\del_3\phi}{f_\pi \sinh\left(\frac{\pi}{4}\mathcal B_3 \right)} \cosh\left(\frac{\pi}{4}\mathcal B_3 \right)\right) \frac{1}2\\
        A_0 &= \frac{\del_3\phi}{f_\pi \sinh\left(\frac{\pi}{4}\mathcal B_3 \right)} \left(\cosh\left(\frac{\mathcal B_3}{2} \arctan{z} \right) - \cosh\left(\frac{\pi}{4}\mathcal B_3 \right)\right) \frac{1}2 + \mu_B .\\
\end{aligned}
\end{equation}
Deriving, we can find the associated field strengths:
\begin{equation}
\begin{aligned}
        A_0 &= \frac{\del_3\phi}{f_\pi \sinh\left(\frac{\pi}{4}\mathcal B_3 \right)} \left(\cosh\left(\frac{\mathcal B_3}{2} \arctan{z} \right) - \cosh\left(\frac{\pi}{4}\mathcal B_3 \right)\right) \frac{\mathbf 1}2 + \mu_B \mathbf 1 \\
        F_{30} &= \frac{\del_3^2\phi}{f_\pi \sinh\left(\frac{\pi}{4}\mathcal B_3 \right)} \left(\cosh\left(\frac{\mathcal B_3}{2} \arctan{z} \right) - \cosh\left(\frac{\pi}{4}\mathcal B_3 \right)\right) \frac{\mathbf 1}2 \\
        F_{z0} &= K^{-1} \frac{\mathcal B_3\del_3\phi}{2f_\pi \sinh\left(\frac{\pi}{4}\mathcal B_3 \right)} \sinh\left(\frac{\mathcal B_3}{2} \arctan{z} \right) \frac{\mathbf 1}2 .\\
\end{aligned}
\end{equation}
Again note that eq.~\eqref{eq:fz3proptoA0} along with the boundary conditions require that $f_{z3}$ non-varying in $x^3$ if not vanishing. This implies that $\del_3^2 \phi = 0$.
The $F_{30}$ is kept as is because it will be use full for the approximation of the massive case \eqref{sec:bulkmn0}.

\subsection{Case: $\mathcal B_3 = 0$ and $m_\pi\neq 0$}
\label{app:solution_mn0B0}
Here we show the derivation of bulk fields for $\mathcal B_3 = 0$ while the mass is left generically non-vanishing.
Furthermore, the non-vanishing mass term leads to a non-local contribution to the first equation of motion.
 As such the equations of motions can be written as the following:
\begin{equation}
    \begin{aligned}
        \partial_3 \left(K F_{z3} \right) &= -\frac{2}{f_\pi^2\pi} \frac{\delta \mathcal L_\mathbf{mass}}{\delta A_z} = -\frac{2}{\pi} m_\pi^2 \sin(f_\pi^{-1} \phi) \frac{\tau^3}{2} \\
        \partial_z \left(K F_{z3} \right) &= 0 \\
        \partial_z \left(K F_{z0} \right) + \partial_3 \left(K^{-1/3} F_{30} \right)  &= 0 .
    \end{aligned}
\end{equation}
The second equation implies that 
\begin{equation}
    F_{z3} = \beta_1(x^3) K^{-1}\,.
\end{equation}
To use the first equation we must use the boundary conditions for $A_3$:
\begin{equation}
    F_{z3} = \del_z A_3 \implies A_3 = \beta_1(x^3) \arctan(z) + \beta_2(x^3)\,.
\end{equation}
The boundary conditions $A_3\vert_{z=\pm\infty} = \mp f_\pi^{-1} \del_3\phi\frac{\tau^3}{2}$ imply
\begin{equation}
\beta_1 = -\frac{1}{\pi f_\pi} \del_3 \phi \tau^3
\end{equation}
and $\beta_2 = 0$.
Therefore the first equation motion implies 
\begin{equation}
\frac{1}{\pi f_\pi} \del_3^2 \phi = \frac{1}{\pi} m_\pi^2 \sin(f_\pi^{-1} \phi)
\end{equation}
where $\phi$ is equal to the ChPT CSL solution
\begin{equation}
    \phi (x^3) = f_{\pi}  \left[\pi \pm 2\mathrm{am}\left( m_\pi k^{-1} x^3, k \right) \right]\,.
\end{equation}
For completeness we remark that because of boundary conditions and the third equation of motion, $A_0 = \mu_B$.

\section{Bulk mass action variation}
\label{app:mass_action_variation}

In this appendix, we will show how 
 the variation of $S_{\text{mass}}$ gives the mass term in eq.~\eqref{eq:masslesspion_eomz}.
First, let's express $U$ in terms of bulk fields:
\begin{equation}
    U = \mathcal P \exp\left(-i \int_{-\infty}^\infty A_z dz\right)
\end{equation}
where $\mathcal P$ is the path ordering operator.
Due to this path ordering, variations must be taken carefully.
Tentatively, we write $\delta U$ as
\begin{equation}
    U + \delta U = \mathcal P \exp\left(-i \int_{-\infty}^\infty (A_z + \delta A_z) dz\right).
\end{equation}
There is an alternative definition of an ordered exponential:
\begin{equation}
    U + \delta U = \prod_z \exp(-iA_zdz -i\delta A_zdz).
\end{equation}
Keeping first order with respect to $\delta A_z$, we can focus the variation at one point $z=z'$
\begin{equation}
    \left(\prod_{z<z'} \exp(-i A_zdz )\right) (-i \delta A_z dz') \left(\prod_{z>z'} \exp(-i A_zdz )\right).
\end{equation}
The total variation involves integrating over all $z'$, resulting in
\begin{equation}
    \delta U = \int dz' \left(\prod_{z<z'} \exp(-i A_zdz )\right) (-i \delta A_z(z')) \left(\prod_{z>z'} \exp(-i A_zdz )\right).
\end{equation}
Now let's consider the variations of $\delta U$ after being traced as in the mass action \eqref{eq:action_pionmass}:
\begin{equation}
    \text{Tr}(\delta U) = \int dz' \text{Tr}\left(\left(\prod_{z<z'} \exp(-i A_zdz )\right) (-i \delta A_z(z')) \left(\prod_{z>z'} \exp(-i A_zdz )\right)\right).
\end{equation}
The permutation property of the trace we can bring the right matrix to the left
\begin{equation}
    \text{Tr}(\delta U) = \int dz' \text{Tr}\left(\left(\prod_{z>z'} \exp(-i A_zdz )\right)\left(\prod_{z<z'} \exp(-i A_zdz )\right) (-i \delta A_z(z')) \right).
\end{equation}
This result implies that we can bring the action to the form where the variation, $\delta A$, is on the right side. 
Thus we can consider $-i \left(\prod_{z>z'} \exp(-i A_zdz )\right)\left(\prod_{z<z'} \exp(-i A_zdz )\right)$ to contribute to the equations of motion by the variation of $A_z$.
A similar argument follows for $\delta U^\dagger$.
Now restricting $A_3$ to only have generators of the commuting subalgebra of $\mathfrak{su}(N_f)$, contribution of the equation of motion is simply $-iU$ from $\delta U$ and $iU^\dagger$ from $\delta U^\dagger$.
Therefore, for our work's setup we can write the variation of the mass action \eqref{eq:action_pionmass} as
\begin{equation}
    \begin{aligned}
        \delta S_{\text{mass}} &= \int d^4 x \frac{m_\pi^2 f_\pi^2}{4} \text{Tr} \left(\delta U + \delta U^{-1} \right)\\
                               &= \int d^4 x dz \frac{m_\pi^2 f_\pi^2}{4} \text{Tr} \left((-iU + iU^\dagger)\delta A_z\right) \\
                               &= \int d^4 x dz \frac{m_\pi^2 f_\pi^2}{4} \text{Tr} \left(-i(U-U^\dagger)\delta A_z\right) \\
                               &= \int d^4 x dz m_\pi^2 f_\pi^2 \text{Tr} \left(\sin(f_\pi^{-1}\phi)\frac{\tau^3}{2}\delta A_z\right). \\
    \end{aligned}
\end{equation}
For the second the last line the gauge condition in tandem with the boundary conditions was used to say that $U$
\begin{equation}
     \frac{\delta \mathcal L_\mathbf{mass}}{\delta A_z} = m_\pi^2 f_\pi^2 \sin(f_\pi^{-1} \phi) \frac{\tau^3}{2}.
\end{equation}
Here, we use the bulk boundary conditions in eq.~\eqref{eq:boundary_conditions}.

\bibliographystyle{jhep}
\bibliography{reference}

\end{document}